%% file: paper.tex
\documentclass[twocolumn,showpacs,prb,superscriptaddress,aps,floatfix]{revtex4-1}

\usepackage{rotating}
\usepackage{amsmath}
\usepackage{amssymb}
\usepackage{color}
\usepackage{graphicx}
\usepackage[active]{srcltx}
\usepackage[sort&compress]{natbib}
\usepackage{amssymb}
\usepackage[dvips]{epsfig}
\usepackage{float}
\usepackage{braket}
\usepackage{slashed}
\usepackage{tikz}
\usetikzlibrary{positioning,shapes,shadows,arrows}
\usepackage{verbatim}
\usepackage{mathtools}
\usepackage{dsfont}

\tikzstyle{element} = [rectangle, rounded corners, minimum width=3cm, minimum height=1cm,text centered, draw=black, fill=yellow!20]
\tikzstyle{line} = [draw, -latex']

\include{alias}

\begin{document}

\title{An unified theory of quantised electrons, phonons and photons out-of-equilibrium: a simplified {\em ab--initio} approach based on the Generalised Baym--Kadanoff ansatz}

\author{Pedro Melo}
\affiliation{\cnr}
\affiliation{\cfc} 

\author{Andrea Marini}
\affiliation{\cnr} 
\affiliation{\etsf} 

\date{\today}
\begin{abstract}
We present a full \emph{ab-inito} description of the coupled out--of--equilibrium dynamics of photons, phonons, and electrons. In the present approach the
quantised nature of the electromagnetic field as well as of the nuclear oscillations is fully taken into account. The result is a set of integro--differential
equations, written on the Keldysh contour, for the Green's functions of electrons, phonons, and photons where the different kind of interactions are merged
together. We then concentrate on the electronic dynamics in order to reduce the problem to a computationally feasible approach. By using the Generalised
Baym--Kadanoff ansatz and the Completed Collision approximation we introduce a series of efficient but controllable approximations. In this way we reduce all
equations to a set of decoupled equations for the density matrix that describe all kind of static and dynamical correlations. The final result is a
coherent, general, and inclusive scheme to calculate several physical quantities: carrier dynamics, transient photo--absorption and light--emission. All of which
include, at the same time, electron--electron, electron--phonon, and electron--photon interaction. We further discuss how all these observables can be easily
calculated within the present scheme using a fully atomistic {\em ab--initio} approach.
\end{abstract}      

\pacs{71.10.-w,78.47.D-,31.15.A-}





\maketitle

\section{Introduction}
\label{intro}
The impressive progresses in ultrafast and ultrastrong laser--pulse technology has paved the way to the modern non--equilibrium\,(NEQ) attosecond
spectroscopies\cite{bgk.2009,spsk.2012,ghlsllk.2013,kc.2014,science.2014}. Unlike conventional spectroscopies, the sample is driven away from equilibrium by a
strong laser pulse (the pump) and then probed with a weaker field (the probe). This second field is used to monitor a wealth of physical properties of the material
in order to disclose the complex properties of the excited system trough the different phases of the real--time evolution\cite{Rossi2002,Krausz2009}. Indeed,
experiments are carried out using pump pulses with frequency in the infrared-ultraviolet range and ultrashort probe pulses (down to a few hundreds of
attoseconds). By varying the delay between the pump and the probe pulses one can monitor the excited-state dynamics in a wide energy and time range.

The elemental processes that are induced by the perturbation with a strong laser field and studied in the present work are schematically represented
in Fig.\ref{fig:1}. 

The external laser pulse first excites a certain density of carriers from the valence to the conduction bands. The duration of this process is directly
controlled by the duration of the laser field. In addition the density of carriers is dictated by the intensity of the laser that also controls the amount of
energy transferred to electrons and holes\cite{Rossi2002,Krausz2009,Axt1998,Kadanoff1962,hj-book,stefanucci,Bonitz1998}.

Already during this first step of the whole dynamics static and coherent correlation effects play a crucial role. As an example, it is well known that the light
absorption process is accompanied by the formation of excitonic states~\cite{Onida2002}. These are bound electron--hole pairs created by the initial laser
excitation. The electron--hole attraction is described by a screened Coulomb interaction. Nevertheless, at this stage the quantisation of the
electromagnetic field is not requested and indeed, in studying simple optical absorption the simpler classic treatment is commonly
adopted\cite{Onida2002,Attaccalite2011}.

After the photo--excitation the carriers will relax by dissipating and transferring energy among themselves and to the lattice. 
During this first step of the dynamics (that can be as long as few pico--seconds)
the ensemble of electrons interact via repeated collisions mediated by the Coulomb\,(e--e) and the electron--phonon\,(e--p) interactions.
Indeed, the complex interaction of the carriers with the lattice imposes the quantisation of the atomic oscillations in the form of phonon modes. These then enter naturally in the
dynamics and must be included in a coherent framework together with excitonic effects. This already represents a first important and challenging aspect in the description
of out--of--equilibrium processes.

\begin{widetext}
The last step in the dynamics of the photo--excited carriers is the recombination with the consequent spontaneous light emission. This process is entirely a
quantistic phenomenon and requires the quantisation of the electromagnetic field. As a consequence, it is evident that a comprehensive description of all phases
of the dynamics following a photo--excitation event require the simultaneous quantistic treatment of electron, phonons and photons. This is well beyond the
state--of--the art and it represents the goal of the present work.
\begin{figure}[H]
\begin{center}
\epsfig{figure=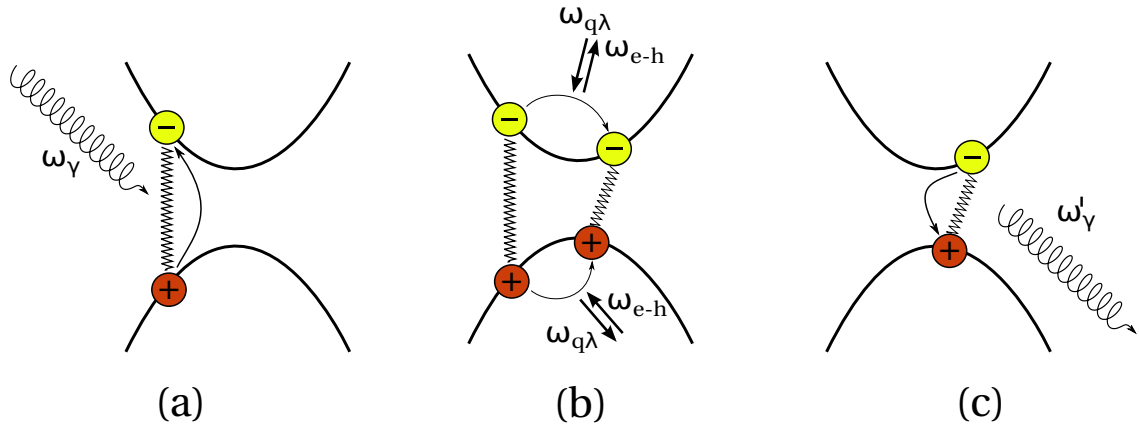,width=15cm}
\caption{\footnotesize{
Schematic representation of the different processes induced by the interaction of a material with a short and intense laser pulse. 
The action of the laser creates electron--hole pairs [process (a)]. These pairs interact trough the screened Coulomb interaction (zig--zag line) creating 
transient excitonic states. After the photo--excitation the carriers undergo repeated collisions with other carriers and with the lattice [process (b)].
During these processes phonons and electron--hole pairs are emitted and/or absorbed. Finally, after pico to microseconds, the excited carriers eventually
relax to the ground state emitting photons [process (c)]. $\go_{\gc}$ and $\go'_{\gc}$ are photon energies, $\go_{\qq\gl}$ is a phonon energy, and
$\go_\mathrm{e-h}$ represents and electron--hole pair energy.
}}
\label{fig:1}
\end{center}
\end{figure}
\end{widetext}

In a typical Pump\&Probe experiment\,(P\&p), a weaker and perturbative second laser pulse is used to probe the system at any time between the processes (a) and
(c) of Fig.\ref{fig:1}. A wealth of time--dependent observables are then measured experimentally. Examples are: the change in the absorption of the probe
induced by the pump (transient absorption)\cite{Koch2006/07//print,Shi2013,Norris1982}, the time--dependent light emission spectrum\cite{Lefebvre2008}, the
time--dependent photo--electron spectrum\cite{Ichibayashi2009,Sangalli2015}.

The key point is that all these observables are inherently connected, as they are produced from the same elemental dynamics of the system, although from
different perspectives. 

From a theoretical point of view, the device of a coherent description of the excited state of the materials has followed several distinct and often fragmented
and uncorrelated paths. The most up--to--date scheme to calculate and predict the ground-- and excited--state properties of a wide range of materials is based
on the merging of Density--Functional--Theory\,(DFT)\cite{R.M.Dreizler1990} with Many--Body Perturbation Theory\,(MBPT)\cite{Onida2002}. DFT is a broadly used
\textit{ab-initio} ground--state theory, that allows to calculate \textit{exactly} the electronic density and total energy without adjustable parameters. The
merging of DFT with perturbation theory gives the so-called Density-Functional Perturbation Theory\cite{Gonze1995,baroni2001}\,(DFPT). The DFPT is a powerful
computational tool for the direct treatment of phonons. However, the DFT computation of excited electronic states properties, like the band-gap energies, is a
known problematic topic~\cite{R.M.Dreizler1990}. As a result, MBPT is nowadays the preferred alternative to DFT for that purpose. It is based on the accurate
treatment of correlation effects by means of the Green's function formalism. MBPT is formally correct and leads to a close agreement with
experiment\cite{VanSchilfgaarde2006}, but is extremely computationally demanding. A natural way to solve this issue is to merge the quick DFT calculation with
the accurate MBPT one. The latter method is often referred to as \textit{ab-initio} Many-Body Perturbation Theory~\cite{Onida2002} (\textit{ai}--MBPT). In this
method, DFT provides a suitable single--particle basis for the MBPT scheme. This methods has been applied successfully, for example, to correct the well--known
band--gap underestimation problem of DFT~\cite{GW_review}.

Therefore, as far as equilibrium properties are concerned, the theoretical and methodological developments have constantly contributed to create a consistent
and efficient environment that can now count on a number of well established standards and codes.

The situation in the out--of--equilibrium case is rather different. Indeed the standard tools of equilibrium MBPT cannot be applied and one has to switch to
more advanced non--equilibrium Green's function\,(NEGF) techniques. From a purely theoretical point of view the NEGF theory has been extensively studied
and reviewed in many books~\cite{Kadanoff1962,hj-book,stefanucci,Bonitz1998}. Nevertheless its development has been mainly confined to simple models or
specific models suitable to interpret specific properties. A merging with DFT, in the NEGF case, is still at the very beginning and an inclusive approach is
lacking. As a main consequence there are not standard numerical tools that can be used even by non--experts or experimentalists to support their observations.
This is the main motivation of the present work.


The NEGF theory is indeed, as far as the electron--electron\,(e--e) interaction is concerned, at an excellent level of development. By using the Keldysh contour
formalism~\cite{d.1984} we can obtain the Baym--Kadanoff equations\,(BKE) that govern the electronic motion. The fundamental ingredient of the BKE
is the self--energy, which embodies all the information on the many--particle interaction. The self--energy can be calculated from single--particle
quantities~\cite{d.1984}. The BKE can also be reduced to a Boltzmann--like equation in the Markovian limit~\cite{d.1984}. 


The e--p interaction has been widely studied both at the and out of the equilibrium. In the first case the MBPT approach has been applied and
reviewed in Ref.\onlinecite{Leeuwen2004a}. In this case the nuclear effects are included by deriving a full set of self--consistent equations for the screened
interaction and the self--energy operator. Also the merging of MBPT and DFT has been studied\cite{PhysRevB.91.224310,SP_2014} and
applied~\cite{Cannuccia2011a,Marini2008,Antonius2014,Giustino2010,Zacharias2015} extensively. In the out--of--equilibrium case the theory is well known as well
and it leads to a combined description of both the electronic and phononic dynamics~\cite{hj-book,Allen1987,DasSarma1990}. In this case the merging with DFT has been only
very recently introduced within the simplified Markovian limit~\cite{Bernardi2014} or in the more general scheme based on the Generalised
Baym--Kadanoff ansatz\,(GBKA) and on the Completed Collision approximation\,(CCA)~\cite{m.2012,Sangalli2015}.

It is crucial to note, at this point, that in all the above cases the electron--photon\,(e--$\gamma$) interaction is neglected and the electromagnetic interaction is treated
classically. As we will see shortly this approach is suitable unless we are not interested in the long--time carrier dynamics and in the transient light
emission spectrum.

Indeed another family of theoretical studies are connected to the bridging of the very general theories based on the NEGF with the actual experiments that are 
performed in a typical P\&p setup. 

The case of the carrier dynamics is the most natural and easy to introduce as it is a simple by product of the BKE. Once that GBKA and the CCA are applied the
equation of motion for the electronic occupations is obtained~\cite{m.2012,Sangalli2015,Bernardi2014,Allen1987,DasSarma1990}.

The time--resolved light absorption of the probe field has been studied within the NEGF formalism in Ref.\onlinecite{Perfetto2015}.
Starting from the BKE and introducing an adiabatic approximation, an equation of motion for the linear response function is derived and shown to reduce to
the well--known Bethe--Salpeter Equation\,(BSE) in the absence of the pump field or when this is weak enough to be possible to apply the
low--intensity regime approximation\,(LIA). Both the LIA and the adiabatic ansatz will be introduced and extensively used in this work. 



In order to study photoluminescence\,(PL) the inclusion of the e--e and e--p interactions is not enough. Indeed light emission is possible only when 
also the electromagnetic field is quantised and it appears as a evolving term in the BKE. As in the e--e and e--p cases also the inclusion
of the e--$\gc$ interaction has been studied~\cite{Henneberger1988,PhysRevLett.86.2451,PhysRevB.58.2064}. In these works 
the theory is bridged with the BSE but the e--p interaction, and therefore the photo--excited carriers relaxation, is neglected. Alternatively 
e--p interaction has been included~\cite{PhysRevLett.86.2451,Hannewald2000,Hannewald2003} but, in this case, neglecting the e--e interaction and, therefore,
the formation of excitonic states.

A field that has been developed in parallel to MBPT is the one based on DFT. In DFT the most used tool is the time--dependent Kohn--Sham
equations~\cite{R.M.Dreizler1990} that have been applied to study out--of--equilibrium regimes~\cite{PhysRevB.62.7998,PhysRevB.85.045134}. DFT provides an
alternative and powerful tool where, however, the description of e--e and e--p interactions is quite problematic. Indeed, DFT has been mainly applied to
low--dimensional systems~\cite{rozzi-nature} or to study very short--time effects~\cite{doi:10.1021/acs.jctc.5b00621} where relaxation, dissipation and
formation of excitonic states can be, within specific approximations, neglected.  Quantised electromagnetic fields has been recently introduced within
DFT~\cite{PhysRevA.90.012508}.


This short introduction proves that the development of a theory of  quantised electrons, phonons and photons out--of--equilibrium is fragmented in several, often
uncorrelated, paths. In addition these approaches often use very different schemes and approximations, often difficult to merge together.
In addition, the complexity of all these theories has prevented their application to realistic materials at a level of accuracy comparable with the
\textit{ai}--MBPT approach. 
This has deprived the theory of its potential predictive power. Also because, in general, P\&p experiments are used to measured changes in specific observables,
which are often very small quantities, therefore making the accuracy of the theory very important.
The merging with DFT has been only recently introduced and it is still at the
initial stages of its development~\cite{m.2012,Attaccalite2011,Bernardi2014,Bernardi2015}.

In this paper we present a full description of an interacting system, thus including electrons, photons and phonons, out--of--equilibrium.
We start from the many--body Hamiltonian and by following a step by step mathematical procedure based on NEGF theory we derive the complete BKE including coherently
the three interactions: e--e, e--p and e--$\gc$. In this way we go beyond the state--of--the--art knowledge preparing the field for the merging with DFT.
Indeed, in order to do so, we discuss key physical regimes (the low--intensity and adiabatic regimes) from which we derive key approximations. By using these
approximations the overall complexity of the equations will be greatly simplified and suitable to be merged with DFT in a computationally feasible approach 
to out--of--equilibrium phenomena.


The structure of the paper is at it follows.  We will first introduce the problem, the Hamiltonian (Sec.\ref{sec2}) and the interaction terms (Sec.\ref{sec2A}).
By extending the seminal works of Ref.\onlinecite{Leeuwen2004a} on the electron--phonon interaction and of
Ref.\onlinecite{Henneberger1988,PhysRevB.58.2064,Pereira1996,d.1984} on the electron--photon interaction, we will introduce in Sec.\ref{sec3A} the equation of
motion for the Green's function, defined in Sec.\ref{sec3}, on the Keldysh contour. We will discuss extensively the longitudinal and transverse response
functions (Sec.\ref{sec3C}) and vertex functions (Sec.\ref{sec3E}). Although we derive a full consistent set of equations for all the different parts entering
the scattering process, here we will focus on the electronic dynamics. The equations of motion relative to the other particles (photons and phonons) will be
introduced in a future work. 
In Sec.\ref{sec4} we will introduce the GW approximation and, in Sec.\ref{sec5}, this
approximation will be used to close the dynamics in the space of the single--time density matrices by means of the GBKA (Sec.\ref{sec5}). In addition we will
investigate further simplifications obtained by treating the memory effects within the CCA (Sec.\ref{sec5C}). Thanks to these crucial approximations we will
arrive to a closed equation, in Sec.\ref{sec6}, that will allow us to derive simple but efficient equations to describe the carrier dynamics (Sec.\ref{sec6B}),
the transient optical absorption (Sec.\ref{sec6C}) and, finally, the light emission spectrum (Sec.\ref{sec6D}). We will conclude this work by rewriting the
theory in the frequency domain by introducing an adiabatic ansatz (Sec.\ref{sec7}) and by discussing the merging with the {\em ab--initio} methods
(Sec.\ref{sec8}).

\section{The bare Hamiltonian and the interaction terms}
\label{sec2}
We start with the non relativistic Hamiltonian of a system of interacting electrons moving under the action of an external
electromagnetic field and of the internal electron--nucleus interaction:
\begin{gather}
\hat{H}=\hat{H}_{0}+ \hat{H}_\mathrm{int},
\label{sec2:eq:full_h}\\
\hat{H}_0=\sum_i h\(\hat{\rr}_i\)+ \hat{T}_\mathrm n +\hat{H}_{\gc}+ \hat{H}_\mathrm{n-n},
\tag{\ref{sec2:eq:full_h}$'$}\\
\hat{H}_\mathrm{int}=\hat{H}_\mathrm{e-e} + \hat{H}_\mathrm{e-n}+\hat{H}_{\mathrm e-\gc}.
\tag{\ref{sec2:eq:full_h}$''$}
\end{gather}
Eq.\eqref{sec2:eq:full_h} includes terms describing the single--particle dynamics ($\hat{T}_e$, $\hat{T}_n$ and $\hat{H}_{\gc}$) and interaction terms 
due to the mutual interaction of electrons, nuclei and photons ($\hat{H}_\mathrm{n-n}$, $\hat{H}_\mathrm{e-n}$, $\hat{H}_\mathrm{e-e}$ and $\hat{H}_{\mathrm{e}-\gc}$).
In the above equation, $h\(\rr\)$ represents the single--particle operator and it is summed over the electronic positions $\rr_i$.

Depending on the choice of the non--interacting part of the Hamiltonian,
$h$ can include, beside the kinetic part, some kind of initial correlation in the form of a mean--field potential. This is an essential
ingredient in the merging of many--body techniques with {\em ab--initio} methods.

The electronic and nuclear kinetic parts are
\begin{gather}
\hat{T}_\mathrm e = -\frac{1}{2}\int \mathrm d^3r\, \hat{\psi}^\dagger(\mathbf r)\nabla^2\hat{\psi}(\mathbf r),
\label{sec2:T}\\
\hat{T}_\mathrm n = -\sum_{\RR} \frac{\nabla_{\RR}^2}{2 M_{\RR}},
\tag{\ref{sec2:T}$'$}
\end{gather}
with $\RR$ the generic position of the nucleus with mass $M_{\RR}$.
In Eq.\eqref{sec2:T}
$\hat{\psi}(\mathbf r)$ and $\hat{\psi}^\dagger(\mathbf r)$ are, respectively, the electron creation and annihilation operators in the Schr\"odinger's picture. 
The spinorial degrees of freedom are not considered here to keep the notation as simple as possible. The extension of the present theory to include their effect 
can be done starting from Pauli's equation and using the minimal coupling transformation. We also use the convention of representing vectors with a bold symbol
(like $\AA$) and tensors using a double arrow over--script (like $\overleftrightarrow{\mathcal D}$). Through this work band indices will be represented by a Latin subscript $(i,j,...)$, while Cartesian directions and branch indices will be denoted by Greek labels $(\ga,\gb,...)$.

$\hat{H}_{\gc}$ is the non-interacting Hamiltonian for the transverse photons
\begin{align}
\hat{H}_{\gamma} = \sum\limits_{\mathbf q, \lambda}\omega_\mathbf q\hat{d}_{\mathbf q,\lambda}^\dagger\hat{d}_{\mathbf q, \lambda},
\label{sec2:eq:h_g}
\end{align}
with $\mathbf q$ and $\omega_\mathbf q$ the photon's momentum and energy, $\lambda$ its polarisation, and $\hat{d}_{\mathbf q,\lambda}^\dagger$ and $\hat{d}_{\mathbf q, \lambda}$ 
the creation and annihilation operators, respectively (also in Schr\"odinger's picture). 

The first group of interaction terms describe the nuclear--nuclear and e--e interactions. Those do not make the different sub--spaces of 
electron, nuclei and photons interact but build up internal (purely electronic and nuclear) correlation effects:
\begin{align}
\hat{H}_\mathrm{e-e} = \frac{1}{2}\int \mathrm d^3r\mathrm d^3r'\, \hat{\psi}^\dagger(\mathbf r)\hat{\psi}^\dagger(\mathbf r')v(\mathbf r - \mathbf r')\hat{\psi}(\mathbf r')\hat{\psi}(\mathbf r),
\label{sec2:eq:h_e-e}
\end{align}
with $v\(\rr-\rr'\)$ the bare Coulomb potential. The nucleus--nucleus interaction reads:
\begin{align}
\tilde{H}_\mathrm{n-n}=\frac{1}{2}\sum_{\RR,\RR'}\nolimits'Z_{\RR} Z_{\RR'}v\(\RR-\RR'\),
\label{sec2:eq:h_n-n}
\end{align}
with $\sum_{ij}\nolimits'=\sum_{i\neq j}$ and $Z_{\RR}$ the nucleus atomic number. 

The e--p interaction, $\hat{H}_\mathrm{e-\gc}$, is given by
\begin{align}
\hat{H}_\mathrm{e-\gamma} = - \frac{1}{c}\int \mathrm d^3r\, \hat{\mathbf{A}}(\mathbf r)\cdot \hat{\mathbf{J}}(\mathbf r) + \frac{1}{2c^2}\int \mathrm d^3r\, \hat{\rho}(\mathbf r)\hat{\mathbf{A}}^2(\mathbf r).
\label{sec2:eq:h_e-g}
\end{align}

Here, the paramagnetic electronic current and the density are defined as
\begin{gather}
\hat{\mathbf{J}}(\mathbf r) = \frac{1}{2\mathrm i} \[\hat{\psi}^\dagger(\mathbf r){\bf\nabla}\hat{\psi}(\rr) - \mathrm{c.c.}\],\label{sec2:eq:j}\\
\hat{\rho}(\mathbf r) = \hat{\psi}^\dagger(\mathbf r)\hat{\psi}(\mathbf r)\tag{\ref{sec2:eq:j}'}.
\end{gather}

We work in the second quantisation formalism and introduce 
a suitable single--particle basis with orthonormal wave--functions $\{\varphi_{i}\(\rr\)\}$. 
Then the creation and 
annihilation field-operator $\hat{\psi}^{\dagger}\(\rr\)$ and $\hat{\psi}\(\rr\)$ for a particle at position $\rr$ in space are expanded according to 
\begin{align}
\hat{\psi}\(\rr\)=\sum_{i}\varphi_{i}\(\rr\)\hat{c}_{i}.
\label{eq:2nd_quant1}
\end{align}
The one--particle density--matrix operator takes the form
\begin{align}
\hat{\gr}(\rr,\rr')=\hat{\psi}^{\dag}(\rr)\hat{\psi}(\rr')=\sum_{ij}\varphi^*_{i}(\rr) \varphi_{j}(\rr') \hat{\gr}_{ji},
\label{eq:2nd_quant2}
\end{align}
with $\hat{\gr}_{ji}=\hat{c}^{\dagger}_{i}\hat{c}_{j}$. 

The first important step is the quantisation of the electromagnetic field that, as it will be clear shortly, will appear as an explicit ingredient of the time evolution. 
We start by rewriting the vector potential of the electromagnetic radiation in terms of photon creation and annihilation operators
\begin{multline}
\hat{\mathbf A}(\mathbf r) = \sum\limits_{\mathbf G,\mathbf q, \lambda}\( \frac{2\pi c^2}{ \omega_{\mathbf{q}+\mathbf G}\Omega}\)^\frac{1}{2} 
\[\hat{d}_{\mathbf q+\mathbf G, \lambda}e^{\mathrm i(\mathbf q+\mathbf G)\cdot\mathbf r} \right.\\
\left.+\hat{d}^\dagger_{\mathbf q+\mathbf G, \lambda}e^{-\mathrm i(\mathbf q+\mathbf G)\cdot\mathbf r}\]\mathbf e_\lambda(\mathbf q + \mathbf G),
\end{multline}
with $\Omega$ being the volume of the lattice and $\mathbf{e}_{\lambda}(\mathbf q+\mathbf G)$ the polarisation vectors orthogonal to the photon's momentum $\mathbf q + \mathbf G$.

We aim at describing any kind of system by using a super--cell approach. This means that the periodic part of the system (if any) is represented 
by a unit cell of volume $\Omega_s$ containing $N$ momenta $\qq$ ($\Omega\equiv N \Omega_s$). This unit cell is periodically repeated 
displaced of a generic vector $\GG$ of the reciprocal lattice. In the case of isolated systems $N=1$ and $\Omega_s$ is chosen large enough to avoid 
spurious interactions. 

The e--p interaction arises from the term $\hat{H}_\mathrm{e-n}$ where the nuclear and electronic densities are coupled via the Coulomb interaction
\begin{align}
\label{sec2:eq:h_ep}
\hat{H}_\mathrm{e-n} = -\int \mathrm d\rr\mathrm d\RR\,\frac{\hat{\rho}(\mathbf r)\hat{N}(\mathbf R)}{|\mathbf r - \mathbf R|}.
\end{align}
$\hat{N}\(\mathbf R\)$ is the nuclear density operator that we take as the counterpart of the electronic case. The actual definition of the nuclear density
is a delicate issue that has been already discussed in Ref.\onlinecite{Leeuwen2004a}.

\subsection{The coupling to the external perturbations}
\label{sec2A}
The Hamiltonian $H$ describes the complete dynamics of the coupled systems of electrons, photons and phonons. In order to rewrite this dynamics in the form of
equations of motion for the corresponding Green's functions we can use two equivalent paths. One is based on the
standard diagrammatic technique~\cite{ALEXANDERL.FETTER1971} which constructs approximations for the different terms of the theory
by using a geometrical and graphical approach.
An alternative approach, that we follow here, is based instead on the equation of motion approach\cite{strinati}.
This methods leads, both in the equilibrium and the out--of--equilibrium regimes, to a closed set of integro--differential equations that at the equilibrium
are known as Hedin's equations~\cite{Hedin19701}.

In order to extend these equations to account for the correlated dynamics of electrons, photons and phonons in an out-of-equilibrium context we start by
discussing some key aspects of the purely electronic case. In the equation of motion approach the Hamiltonian is perturbed with a fictitious time--dependent term 
\begin{align}
\hat{H}\(t\)=\hat{H}+\hat{H}_\mathrm{ext}\(t\).
\label{sec2:eq:h_t}
\end{align}
In the original derivation of Hedin's equations\cite{strinati} $\hat{H}_\mathrm{ext}\(t\)$ describes the coupling of the electronic charge with an external fictitious field $\phi_\mathrm{ext}\(\rr,t\)$
that, at the end of the derivation, is sent to zero:
\begin{align}
\hat{H}_\mathrm{ext}(t) = \int \mathrm d^3r\, \phi_\mathrm{ext}\(\rr,t\)\hat{\rho}\(\mathbf r, t\).
\label{sec2:eq:h_ext_1}
\end{align}
Now the problem is that this perturbation cannot be used in the present case where we want to describe a quantised electromagnetic 
field. The reason is that the vector potential is now quantised and
it cannot be sent to zero at the end of the calculations. This is connected to the well-known existence of a 
vacuum energy of the electromagnetic field.

To solve this problem we follow a different path. We notice that the external potential $\phi_\mathrm{ext}\(\rr,t\)$ is solution of the Poisson equation
\begin{align}
 \nabla^2 \phi_\mathrm{ext}\(\rr,t\) = -4\pi\gr_\mathrm{ext}\(\mathbf r,t\).
\label{sec2:eq:poisson}
\end{align}
The solution of Eq.\eqref{sec2:eq:poisson} can be rewritten in integral form
\begin{align}
\phi_\mathrm{ext}\(\rr,t\)=\int \mathrm d^3r'\, v\(\rr-\rr'\) \gr_\mathrm{ext}\(\rr',t\),
\label{sec2:eq:poisson_solution}
\end{align}
that, plugged into Eq.\eqref{sec2:eq:h_ext_1}, yields 
\begin{align}
\hat{H}_\mathrm{ext}(t) = \int \mathrm d^3r\,\, \hat\phi\(\rr\)\rho_\mathrm{ext}\(\mathbf r, t\),
\label{sec2:eq:h_ext_2}
\end{align}
with $\gr_\mathrm{ext}$ the inhomogeneous part of Eq.\eqref{sec2:eq:poisson}.

By comparing Eq.\eqref{sec2:eq:h_ext_1} with Eq.\eqref{sec2:eq:h_ext_2} we notice that, now, the potential can be quantised and the external charge sent to zero. As it will be clear in 
the following, these two procedures are equivalent, as far as the solely electronic limit is concerned.

Another important aspect is that we can now couple the external test charge to the potential generated by both the nuclear and the electronic charges:
\begin{multline}
\hat\phi\(\rr,t\)=\int\mathrm d^3r'\, v\(\rr-\rr'\) \[\hat{\gr}\(\rr'\)-\hat N(\rr')\]\\
=\int\mathrm d^3r'\, v\(\rr-\rr'\) \hat n(\rr') ,
\end{multline} 
where $\hat n = \hat\gr - \hat N$ is the total internal density of the system, since now we have to account for the quantised nuclei as well.
By following Ref.\onlinecite{Leeuwen2004a} we introduce an external
perturbation of the nuclear density, $N_\mathrm{ext}\(\RR,t\)$, that is coupled, via Eq.\eqref{sec2:eq:poisson_solution}, to the electric potential generated by
the nuclei, $\hat{V}_\mathrm n\(\mathbf R,t\)$. 

A similar procedure can be applied to the perturbation induced by an external vector potential by using the equation of motion
\begin{align}
\(\frac{1}{c^2}\partial^2_{t} - \nabla^2\)\mathbf{A}_\mathrm{ext}\(\rr,t\)= \frac{4\pi}{c}\mathbf J_\mathrm{ext}\(\rr,t\).
\label{sec2:eq:eom_A}
\end{align}
In this way the total interaction part of the Hamiltonian looks like
\begin{widetext}
\begin{align}
\hat{H}_\mathrm{ext}(t) = \int \mathrm d^3r\, \hat\phi(\mathbf r,t)\rho_\mathrm{ext}(\mathbf r, t) - 
\frac{1}{c}\int \mathrm d^3r\, \hat{\mathbf{A}}(\mathbf r)\cdot \mathbf{J}_\mathrm{ext}(\mathbf r,t) 
 -\int \mathrm d^3R\, \hat{V}_\mathrm n(\mathbf R)N_\mathrm{ext}(\mathbf R,t).
\label{sec2:eq:h_ext}
\end{align}
\end{widetext}
The last term in Eq.\eqref{sec2:eq:h_ext} describes the coupling with the nuclear motion. 
In Eq.\eqref{sec2:eq:h_ext} we have neglected the second term on the right hand side of Eq.\e{sec2:eq:h_e-g}. The reason is that this term 
is multiplied by $1/c^2$ and therefore, for a given order in the perturbative expansion it leads to correction much smaller then the ones
induced by the $\AA\cdot\JJ$ term. From a diagrammatic point of view the $\gr\AA^2$ term is responsible of a series of diagrams well--known and studied
in the e--p problem~\cite{PhysRevB.91.224310}, where they arise from the second--order e--p interaction. In the e--$\gc$ case, 
however, these diagrams can be safely neglected.

\section{Green's functions and equations of motion on the Keldysh contour}
\label{sec3}
\begin{figure}[H]
\begin{center}
\epsfig{figure=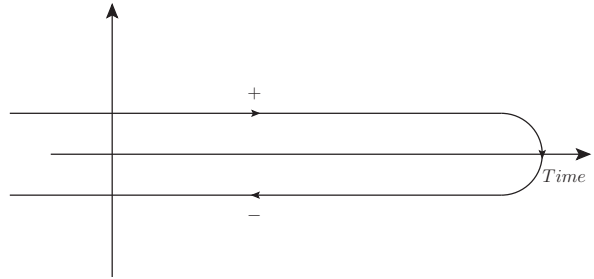,width=8cm}
\caption{\footnotesize{The Keldysh contour. Any time index runs on the contour which defines a natural time ordering that only in the upper branch is equivalent
to the ordering in the standard real--time axis.
}}
\label{fig:2}
\end{center}
\end{figure}

The time evolution in non--equilibrium processes is 
more complicated than for equilibrium systems because it is not guaranteed that after an arbitrarily long enough time the system will return to the ground state. 
The extension of Hedin's equations to an out--of--equilibrium system can be done with recourse to a contour in the complex plane known as the Keldysh
contour~\cite{hj-book}. The contour runs as shown in Fig.~\ref{fig:2} with the upper (positive) branch running from an instant $t_0$ where the
system is assumed to be at equilibrium to an unknown state. The system is brought back to equilibrium via the lower (negative) branch and it is this
mathematical description for the time evolution which permits us to have all the necessary rules to reconstruct an equation system similar to the original one
of Hedin's~\cite{hj-book}.

Therefore, the Green's function $G\(1,2\)$ for non--equilibrium processes is defined on the Keldysh contour~\cite{stefanucci}:
\begin{align}
\label{sec3:eq:g_contour}
G\(1_\ga,2_\gb\) = -\mathrm i\beta\frac{\mathrm{Tr}\left\{\rho_0\mathrm T_C\[\hat S_C \hat\psi_I\(1_\ga\)\hat\psi_I^\dagger\(2_\gb\) \]\right\}}{\mathrm{Tr}\left\{\rho_0\hat S_C\right\}},
\end{align}
where the subscript $I$ indicates that operators are in the interaction picture and subscripts $\alpha,\,\beta = \pm 1$ indicate the branch on the Keldysh contour where the time argument of the respective operator 
is located. The operator $\mathrm T_C$ is the contour time ordering operator.
In Eq.\eqref{sec3:eq:g_contour} we introduce the compact notation $1 \equiv \(\mathbf r_1, t_1\)$. Note that the time $t$ runs on the Keldysh contour.

The transition from the Heisenberg picture (where the electron operators are usually defined) to the interaction one is done via the following expression
\begin{align}
\label{sec3:eq:O}
\hat{O}\(t_\ga\) = \hat{S}\(-\infty,t_\ga\)\hat{O}_I\(t_\ga\)\hat{S}\(t_\ga,-\infty\),
\end{align}
where $\hat{O}$ is a generic operator and
\begin{align}
\label{seq3:eq:s_a}
\hat S\(t_\ga,t'_\ga\) = \mathrm T_\alpha \mathrm{exp}\[- \mathrm i \ga \int_{t'_\ga}^{t_\ga} \mathrm d\tau\, \hat H_{\mathrm{ext},I}^\alpha (\tau)\].
\end{align}
Note that, in Eq.\eqref{seq3:eq:s_a}, the time arguments lie on a single time branch of the Keldysh contour. This implies that we can introduce a branch specific
evolution operator, $\hat S_\ga\(t,t'\)\equiv \hat S\(t_\ga,t'_\ga\)$. With this definition the overall time evolution operator, $\hat{S}_C$, entering Eq.\eqref{sec3:eq:g_contour}
can be rewritten as
\begin{align}
\label{seq3:eq:s_c}
\hat S_C = \hat S_-(-\infty,\infty)\hat S_+(\infty, -\infty),
\end{align}
and we have that the expectation value of $\hat{O}$ taken on the contour is
\begin{align}
\braket{\hat O\(1\)}_C = \frac{\mathrm{Tr}\left\{\rho_0\mathrm T_C\[\hat S_C \hat O_I\(1\)\]\right\}}{\mathrm{Tr}\left\{\rho_0\hat S_C\right\}}.
\end{align}
The structure of Eq.\eqref{sec3:eq:g_contour} with respect to the position of the time arguments on the Keldysh Contour defines the different kind of Green's functions:
\begin{gather}
G\(1_+,2_+\)\equiv G_{c}\(1,2\),
\label{sec3:eq:G_matrix}\\
G\(1_-,2_+\)\equiv G^{<}\(1,2\),
\tag{\ref{sec3:eq:G_matrix}$'$}\\
G\(1_+,2_-\)\equiv G^{>}\(1,2\),
\tag{\ref{sec3:eq:G_matrix}$''$}\\
G\(1_-,2_-\)\equiv G_{\tilde{c}}\(1,2\).
\tag{\ref{sec3:eq:G_matrix}$'''$}
\end{gather}
In Eq.\eqref{sec3:eq:G_matrix} $G_c$ is the time-ordered or causal Green's function, $G^<$ and $G^>$ are the lesser and greater Green's functions, and $G_\text{\~c}$ is the anti-time-ordered Green's function.

\subsection{Equation of motion for the Green's function on the Keldysh contour}
\label{sec3A}
In order to obtain the equation of motion for $G$ we start from $\mathrm i\partial_{t_1}G\(1_\ga,2_\gb\)$. From Eq.\eqref{sec3:eq:g_contour}
and Eq.\eqref{sec3:eq:G_matrix} we get 
\begin{widetext}
\begin{multline}
\label{sec3:eq:g_eom}
\partial_{t_1}\mathrm T_C\[\hat S_C \hat\psi_I\(1_\alpha\)\hat\psi_I^\dagger\(2_\beta\)\] = \delta\(1_\ga-2_\gb\)\mathrm T_C\[\hat S_C \] 
+\mathrm T_C\[\hat S_C \partial_{t_1}\hat\psi_I\(1_\alpha\)\hat\psi_I^\dagger\(2_\beta\)\] \\
+\mathrm i\alpha\mathrm T_C[\hat S_C\hat{H}_\mathrm{ext}\(t_{1\ga}\) \hat\psi_I\(1_\alpha\)\hat\psi_I^\dagger\(2_\beta\)] - 
\mathrm i\alpha\mathrm T_C[\hat S_C \hat\psi_I\(1_\alpha\)\hat{H}_\mathrm{ext}\(t_{1\ga}\)\hat\psi_I^\dagger\(2_\beta\)].
\end{multline}
In the second term of the r.h.s. of Eq.\eqref{sec3:eq:g_eom} we have the time derivative of the electron annihilation operator that is, in the interaction picture,
\begin{align}
\label{sec3:eq:eom_psi}
\mathrm i\partial_{t_1}\hat{\psi}_I(1) = \[\hat{\psi}_I(1),\hat H_I(t_1)\] 
= \[\hat{h}\(1\) - \frac{\mathrm i}{c}\hat{\mathbf A}(1)\cdot\nabla_1 + \hat\phi(1)\]\hat\psi_I(1).
\end{align}
$\hat H_I$ is the representation, in the interaction picture, of $\hat{H}_\mathrm{ext}\(t\)$, defined in Eq.\eqref{sec2:eq:h_t}, while the third and fourth terms in Eq.\eqref{sec3:eq:g_eom} give 
\begin{align}
\label{sec3:eq:com1}
\mathrm i\alpha\mathrm T_C[\hat S_C\hat{H}_\mathrm{ext}(t_1) \hat\psi_I\(1_\alpha\)\hat\psi_I^\dagger\(2_\beta\)] 
-\mathrm i\alpha\mathrm T_C[\hat S_C \hat\psi_I\(1_\alpha\)\hat{H}_\mathrm{ext}(t_1)\hat\psi_I^\dagger\(2_\beta\)] = 
-\mathrm i \ga \phi_\mathrm{ext}(1)\mathrm T_c[\hat S_c\hat\psi(1_\ga)\hat\psi^\dagger(2_\gb)].
\end{align}
Eqs.(\ref{sec3:eq:g_eom}--\ref{sec3:eq:com1}), finally, lead to the following equation of motion for $G$ 
\begin{align}
\label{sec3:eq:eom_G}
\[\mathrm i \partial _{t_1} - \hat{h}\(1\)- \frac{\mathrm i}{c}\braket{\hat{\mathbf{A}}\(1_\ga\)}_C\cdot \nabla_1-U\(1_\ga\) \]G\(1_\ga,2_\gb\) 
=\delta\(1_\ga-2_\gb\) + \left.\mathrm i\frac{\delta G\(1_\ga,2_\gb\)}{\delta \rho_{\mathrm{ext}}\(3_\gc\)}\right|_{3_\gc=1_\ga} - 
\left.\nabla_1\cdot\frac{\delta G\(1_\ga,2_\gb\)}{\delta \mathbf J_{\mathrm{ext}}\(3_\gc\)}\right|_{3_\gc=1_\ga}.
\end{align}
\end{widetext}
In Eq.\eqref{sec3:eq:eom_G} we defined the total potential 
\begin{align}
\label{sec3:eq:u}
U\(1\) = \phi_\mathrm{ext}\(1\) + \braket{\hat{\phi}\(1\)}_C,
\end{align}
and used the following identities for the functional derivatives
\begin{multline}
\label{sec3:eq:deltaG/deltaR}
\frac{\delta G\(1_\ga,2_\gb\)}{\delta \rho_{\mathrm{ext}}\(3_\gc\)} = \mathrm i \gamma\braket{\hat\phi_\gamma\(3_\gc\)}_C G\(1_\ga,2_\gb\) \\
-\gb\gamma\braket{\hat\phi\(3_\gc\)\hat\psi\(1_\ga\)\hat\psi^\dagger\(2_\gb\)}_C,
\end{multline}
and
\begin{multline}
\label{sec3:eq:deltaG/deltaJ}
\frac{\delta G \(1_\ga,2_\gb\)}{\delta \mathbf J_{\mathrm{ext}}\(3_\gc\)} = \mathrm i \frac{\gamma}{c}\braket{\hat{\mathbf A}\(3_\gc\)}_CG\(1_\ga,2_\gb\) \\
+\frac{\gb\gamma}{c}\braket{\hat{\mathbf A}\(3_\gc\)\hat\psi\(1_\ga\)\hat\psi^\dagger\(2_\gb\)}_C.
\end{multline}

Eq.\eqref{sec3:eq:eom_G} can be also used to define the non--interacting Green's function in order to rewrite it as Dyson equation or, alternatively, as 
BKE. Indeed we start by noticing that $G_0$ is the solution of Eq.\eqref{sec3:eq:eom_G} when $\frac{\gd G}{\gd \rho_\mathrm{ext}}=\frac{\gd G}{\gd\JJ_\mathrm{ext}}=0$:
\begin{align}
\label{sec3:eq:eom_Go}
\[\mathrm i \partial _{t_1} - h_\mathrm{ext}\(1_\ga\) \]G_0\(1_\ga,2_\gb\) =\delta\(1_\ga-2_\gb\),
\end{align}
with 
\begin{align}
\label{sec3:eq:eom_ho}
h_\mathrm{ext}\(1\)= h\(1\)- \frac{\mathrm i}{c}\braket{\hat{\mathbf{A}}\(1\)}_C\cdot \nabla_1-U\(1\).
\end{align}
Eq.\eqref{sec3:eq:eom_Go} and Eq.\eqref{sec3:eq:eom_ho} implies that
\begin{align}
\label{sec3:eq:eom_Go2}
G^{-1}_0\(1_\ga,2_\gb\)=\[\mathrm i \partial _{t_1} - h_\mathrm{ext}\(1_\ga\) \]\delta\(1_\ga-2_\gb\).
\end{align}
By using Eq.\eqref{sec3:eq:eom_Go2} we can further rewrite Eq.\eqref{sec3:eq:eom_G} in terms of the bare and of the fully interacting Green's functions. This
is, clearly, a step towards the final form of the Dyson--like equation:
\begin{multline}
\label{sec3:eq:dys1}
G\(1_\ga,2_\gb\)=G_0\(1_\ga,2_\gb\)+G_0\(1_\ga,4_\gs\)\times\\\times\[ 
\left.\mathrm i\frac{\delta G\(4_\gs,2_\gb\)}{\delta \rho_{\mathrm{ext}}\(3_\gc\)}\right|_{3_\gc=4_\gs} - 
\left.\nabla_1\cdot\frac{\delta G\(4_\gs,2_\gb\)}{\delta \mathbf J_{\mathrm{ext}}\(3_\gc\)}\right|_{3_\gc=4_\gs}\].
\end{multline}
The electromagnetic field and the nuclei are now quantised, so the entire theoretical scheme is closed only when the equations of motion for their propagators are introduced. Indeed it is
interesting to note that all different effects induced by the mutual interactions (e--e, e--p, e--$\gc$) will mix together in the dressing of the total fields connected to the external sources.
This dressing will involve the ensemble of electrons, photons, and phonons via oscillations (collective and not) described by the corresponding response functions.

\begin{widetext}
Indeed, as it will be clear shortly, at difference with the well known purely electronic case new response and vertex functions must be introduced.
\begin{figure}
\begin{tikzpicture}[node distance=2.5cm,auto]
\node [element] (box1){ 
  \begin{minipage}{0.202\textwidth}
    \begin{align}
\mathrm i \partial _{t_1} G\(1_\ga,2_\gb\) \nonumber
    \end{align}
\flushright (Sec. III.A)
  \end{minipage}
};
\node [element, below of=box1, left=0.4cm of box1 ] (box2l){ 
  \begin{minipage}{0.202\textwidth}
   \begin{align}
\frac{\delta G\(4_\gs,2_\gb\)}{\delta \rho_{\mathrm{ext}\(3\)}} \nonumber
  \end{align}
 \end{minipage}
};
\node [element, below of=box1, right=0.1cm of box1 ] (box2r){ 
 \begin{minipage}{0.202\textwidth}
  \begin{align}
   \frac{\delta G\(4_\gs,2_\gb\)}{\delta \mathbf J_{\mathrm{ext}}\(3_\gc\)} \nonumber
  \end{align}
 \end{minipage}
};
\node [element, below of=box2l ] (box3l){ 
  \begin{minipage}{0.202\textwidth}
    \begin{gather}
     U\(1\) = \phi_\mathrm{ext}\(1\) + \braket{\hat{\phi}\(1\)}_C\nonumber\\
     W\(1,2\) = \frac{\delta U\(1\)}{\delta\rho_\mathrm{ext}\(2\)}\nonumber
    \end{gather}
\flushright (Sec. III.C.2)
  \end{minipage}
};
\node [element, below of=box2r ] (box3r){ 
  \begin{minipage}{0.202\textwidth}
    \begin{align}
\overleftrightarrow{\mathcal D}\(1,2\) =-\frac{c}{4\pi}\frac{\delta \braket{\hat{\mathbf{A}}\(1\)}_C}{\delta \mathbf J_\mathrm{ext}\(2\)}\nonumber
    \end{align}
\flushright (Sec. III.C.1)
  \end{minipage}
};
\node [element, below of=box3l ] (box4l){ 
  \begin{minipage}{0.202\textwidth}
    \begin{align}
\gamma\(1,2,3\) = -\frac{\delta G^{-1}\(1,2\)}{\delta U\(3\)}\nonumber
    \end{align}
\flushright (Sec.III.D)
  \end{minipage}
};
\node [element, below of=box3r ] (box4r){ 
  \begin{minipage}{0.202\textwidth}
    \begin{align}
{\mathbf\Gamma}\(1,2,3\)= -\frac{4\pi}{c}\frac{\delta G^{-1}\(1,2\)}{\delta \braket{\hat{\mathbf A}\(3\)}}\nonumber
    \end{align}
\flushright (Sec.III.D)
  \end{minipage}
};
\node [element, below=8.0cm of box1] (box5){ 
  \begin{minipage}{0.202\textwidth}
    \begin{align}
\left.\Sigma(1,2) = \mathrm i\[ G(1,3)\gamma(3,2,4)W(4,1^+) +\sum\limits_{\ga,\gb=1}^3\Pi_\ga(1,1')G(1,3)\Gamma_\gb(3,2,4)\mathcal D_{\gb\ga}(4,1')\]\right|_{1'=1}\nonumber
    \end{align}
\flushright (Sec.III.E)
  \end{minipage}
};
\node[below=0.3cm of box4r](d1) {};
\node[below=0.35cm of box4l](d2) {};
\path [line] (box1) -| (box2l);
\path [line] (box1) -| (box2r);
\path [line] (box2l) -- (box3l);
\path [line] (box2r) -- (box3r);
\path [line] (box3l) -- (box4l);
\path [line] (box3r) -- (box4r);
\path [line] (box4r) -- (d1);
\path [line] (box4l) -- (d2);
\end{tikzpicture}
\caption{\footnotesize{ Schematic representation of the mathematical procedure followed to get the Dyson equation within the framework of the Hedin's equations.
The initial derivative is split in longitudinal (left) and transverse (right) contributions.
}}
\label{fig:eq_flow}
\end{figure}
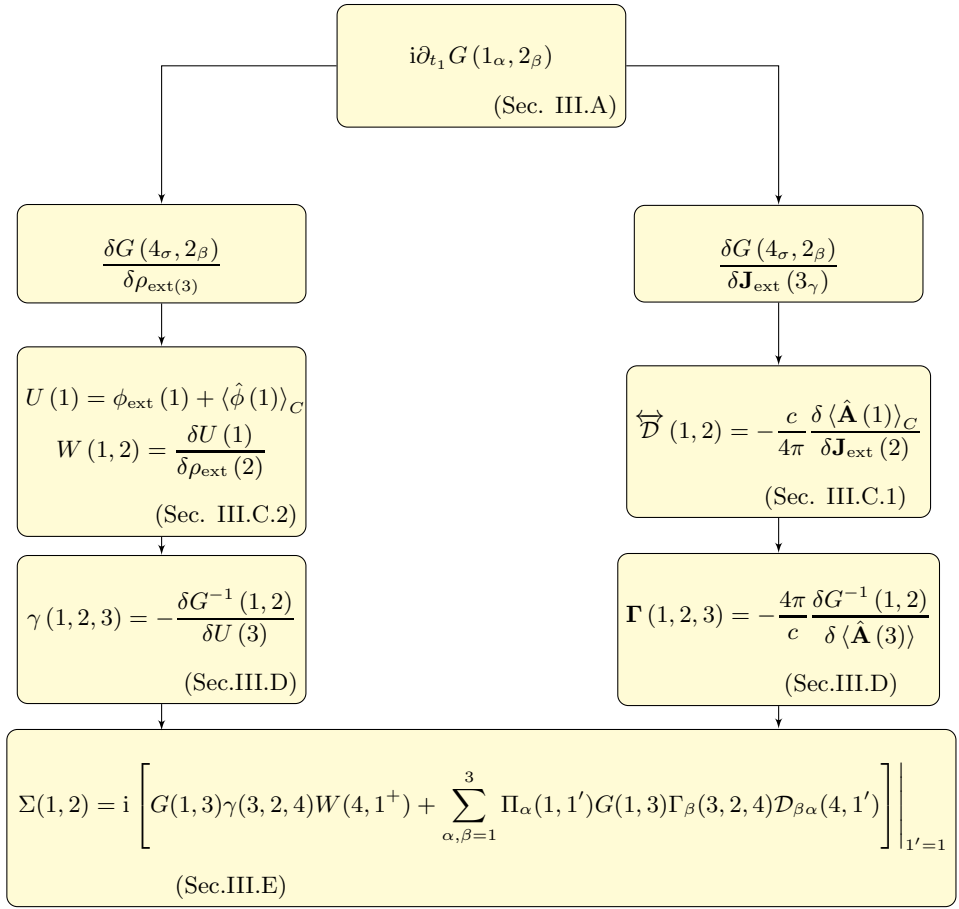
\end{widetext}

\subsection{Equation of motion for the electromagnetic potentials}
\label{sec3B}
The equation of motion for the fields are obtained by taking the macroscopic average of $U$ and 
$\braket{\hat{\mathbf A}}$.
Indeed, from classical electrodynamics, we know that the scalar potential $U$ is the solution of the Poisson equation
\begin{align}
\label{sec2:eq:V_n}
\nabla^2 U(1) = -4\pi[\rho_\mathrm{ext}(1) + \braket{\hat n(1)}_C],
\end{align}
while the expectation value of the vector potential $\braket{\hat{\mathbf A}}$ satisfies the equation of motion
\begin{multline}
\label{sec3:eq:eom_A}
\(\frac{1}{c^2}\partial^2_{t_1} - \nabla_1^2\)\braket{\hat{\mathbf{A}}\(1\)}_C = \frac{4\pi}{c}\mathbf J_\mathrm{tot}^{\perp}\(1\) \\
= \frac{4\pi}{c}\int \mathrm d2\, \overleftrightarrow\delta^\perp\(1,2\)\mathbf J_\mathrm{tot}\(2\).
\end{multline}
The $\perp$ superscript means that only the transverse part of the current enters in the Eq.\eqref{sec3:eq:eom_A}. 
This is obtained from the total current vector using the transverse delta function
\begin{multline}
\delta^\perp_{\ga\gb}\(1,2\) = \delta\(t_1-t_2\)\[\delta(\mathbf r_1-\mathbf r_2)\delta_{\ga\gb} \right.\\
\left.+\frac{1}{4\pi}\frac{\partial}{\partial r_{1,\ga}}\(\frac{1}{|\mathbf r_1 - \mathbf r_2|}\frac{\partial}{\partial r_{2,\gb}}\)\].
\end{multline}

\subsection{Response functions, Vertex functions and Self--energies}
\label{sec3C}
The equation of motion for $G$, Eq.\eqref{sec3:eq:eom_G}, is still written in an obscure way and the different physical ingredients describing the complex many--body
dynamics are hidden inside the functional derivatives $\[\frac{\delta G\(4_\gs,2_\gb\)}{\delta \rho_{\mathrm{ext}\(3\)}}\right.$ and
 $\left.\frac{\delta G\(4_\gs,2_\gb\)}{\delta \mathbf J_{\mathrm{ext}}\(3_\gc\)}\]$. 
In this section we investigate the structure of these derivatives by introducing the longitudinal and transverse response function and vertex functions.

We start by noticing that, as a consequence of the Green's function definition, it follows that
\begin{widetext}
\begin{align}
\label{sec3:eq:dgg}
G\(1_\ga,2_\gb\)G^{-1}\(2_\gb,3_\gc\) = \delta\(1_\ga-2_\gb\)
\Rightarrow \frac{\delta G\(1_\ga,2_\gb\)}{\delta h\(3_\gc\)} = 
-G\(1_\ga,4_\eta\)\[\frac{\delta G^{-1}\(4_\eta,5_\xi\)}{\delta h\(3_\gc\)}\]
G\(5_\xi,2_\gb\),
\end{align}
\end{widetext}
for any well behaved function $h$. In Eq.\e{sec3:eq:dgg} we have introduced the {\em right} $G^{-1}$ function that acts on the right arguments of $G$. Note that
an alternative formulation can be introduced by using the {\em left}  $G^{-1}$ defined in such a way that 
$G^{-1}\(1_\ga,2_\gb\)G\(2_\gb,3_\gc\) = \delta\(1_\ga-2_\gb\)$~\cite{strinati}.

We can now apply a second rule that states that:
\begin{align}
\label{sec3:eq:dgg2}
\frac{\delta G^{-1}\(4_\eta,5_\xi\)}{\delta h\(3_\gc\)} = \frac{\delta G^{-1}\(4_\eta,5_\xi\)}{\delta g\(6_\sigma\)}\frac{\delta g\(6_\sigma\)}{\delta h\(3_\gc\)} .
\end{align}
From Eq.\eqref{sec3:eq:dgg} and Eq.\eqref{sec3:eq:dgg2} we get that, by using $h=\gr$ and $g=U$ (see Eq.\eqref{sec3:eq:u})
\begin{multline}
\label{sec3:eq:deltaG/R}
\left.\frac{\delta G\(1_\ga,2_\gb\)}{\delta \rho_{\mathrm{ext}}\(3_\gc\)}\right|_{3_\gc=1_\ga}=G\(1_\ga,4_\eta\)\gc\(4_\eta,5_\xi,6_\sigma \)\times \\ W\(6_\sigma,1_\ga\)G\(5_\xi,2_\gb\).
\end{multline}
With Eq.\eqref{sec3:eq:deltaG/R} we can now introduce the scalar (longitudinal) part of the total self--energy operator,
$\Sigma_\mathrm{long}$, defined as:
\begin{align}
\label{sec3:eq:sigma_1}
\mathrm i\left.\frac{\delta G\(1_\ga,2_\gb\)}{\delta \rho_{\mathrm{ext}}\(3_\gc\)}\right|_{3_\gc=1_\ga}=\Sigma_\mathrm{long}\(1_\ga,3_\gc\) G\(3_\gc,2_\gb\),
\end{align}
and, from Eq.\eqref{sec3:eq:deltaG/R}, it follows that
\begin{align}
\label{sec3:eq:sigma_2}
\Sigma_\mathrm{long}\(1_\ga,2_\gc\)\equiv G\(1_\ga,3_\gc\) \gc\(3_\gc,2_\gb,4_\sigma \)W\(4_\sigma,1_\ga\).
\end{align}
The definition of the self--energy operator naturally introduces scalar response function
\begin{align}
\label{sec3:eq:W}
W\(1,2\) = \frac{\delta U\(1\)}{\delta\rho_\mathrm{ext}\(2\)},
\end{align}
and the longitudinal vertex function $\gc\(1,2,3\)$ 
\begin{align}
\label{sec3:eq:gamma_long}
\gamma\(1,2,3\) = -\frac{\delta G^{-1}\(1,2\)}{\delta U\(3\)}.
\end{align}

A similar procedure can be applied to the other derivative that appears in Eq.\eqref{sec3:eq:eom_G}, $\frac{\delta G}{\delta \mathbf J_{\mathrm{ext}}}$. At difference with 
the charge derivative this new term is a vector and we use Greek symbols ($\ga,\gb,...$) to label its Cartesian components. By applying the following substitutions
to Eq.\eqref{sec3:eq:dgg} and Eq.\eqref{sec3:eq:dgg2}
\begin{gather}
\label{sec3:tensorial_changes}
\frac{\gd}{\gd \gr_\mathrm{ext}}\Rightarrow \frac{\gd}{\gd J_\mathrm{ext,\ga}},\\
\frac{\gd}{\gd U}\Rightarrow \frac{\gd}{\gd A_{\ga}},
\end{gather}
we get that
\begin{widetext}
\begin{align}
\label{sec3:eq:deltaG/J}
\left. \frac{\delta G\(1_\ga,2_\gb\)}{\gd J_{\mathrm{ext},\lambda}\(3_\gc\)}\right|_{3_\gc=1_\ga}=
-\sum_{\theta=1}^3 G\(1_\ga,4_{\eta}\)\frac{\gd G^{-1}\(4_\eta,5_\xi\)}{\delta \braket{\hat{A}_\theta\(6_\gs\)}_C}\frac{\gd \braket{\hat{A}_\theta\(6_\gs\)}_C}{\gd J_\mathrm{ext,\lambda}\(1_\ga\)}G\(5_\xi,2_\gb\)
\end{align}
Eq.\eqref{sec3:eq:deltaG/J} allows to introduce the second important term in the total self--energy corresponding to its transverse contribution, $\Sigma_\mathrm{trans}$:
\begin{align}
\label{sec3:eq:sigma_trans}
\Sigma_\mathrm{trans}\(1_\ga,2_\gb\)=-
\nabla_i \left. \frac{\delta G\(1_\ga,2_\gb\)}{\gd J_{\mathrm{ext},\lambda}\(3_\gc\)}\right|_{3_\gc=1_\ga}= 
\mathrm i \left.\Pi_i\(1,3\) \sum\limits_{\theta=1}^3G\(1_\ga,4_{\eta}\) \Gamma_\theta\(4_\eta,5_\xi,6_\gs\) \mathcal D_{\theta\lambda}\(6_\gs,3_\gb\)G\(5_\xi,2_\gb\)\right|_{3_\gc=1_\ga}.
\end{align}
\end{widetext}
where 
\begin{align}
\label{sec3:eq:Pi}
\mathbf{\Pi}\(1,2\) =-\frac{\mathrm i}{2}\(\nabla_1 - \nabla_2\) = \mathbf \Pi\(1\) + \mathbf\Pi^*\(2\).
\end{align}
The transverse self--energy, Eq.\eqref{sec3:eq:sigma_trans} further introduces two functions. The transverse photon propagator, which we will call 
$\overleftrightarrow{\mathcal D}(1,2)$
\begin{align}
\label{sec2:eq:trans_prop}
\overleftrightarrow{\mathcal D} \(1,2\) =-\frac{c}{4\pi}\frac{\delta \braket{\hat{\mathbf{A}}\(1\)}_C}{\delta \mathbf J_\mathrm{ext}\(2\)},
\end{align}
and the transverse vertex (vectorial) function $\mathbf{\Gamma}$
\begin{align}
\label{sec2:eq:trans_vertex}
{\mathbf\Gamma}\(1,2,3\)= -\frac{4\pi}{c}\frac{\delta G^{-1}\(1,2\)}{\delta \braket{\hat{\mathbf A}\(3\)}}.
\end{align}
We notice now that the longitudinal screened potential and vertex are connected, via Eq.\eqref{sec2:eq:V_n}, to the {\em total} density. This includes both an electronic
and a nuclear component. It is then clear that the nuclear motion will enter directly in the electronic dynamics via the longitudinal components. 

This sharp separation between longitudinal and transverse components of the theory is possible thanks to the use of the Coulomb's gauge. This allows 
the electromagnetic field to be separated into two independent parts and the entire theory follows the same structure. As we will be seeing shortly, the final
expression for the self--energy will also be split in a longitudinal and in a transverse part.

\begin{widetext}
The final step of this section is obtained by using Eq.\eqref{sec3:eq:sigma_trans} and Eq.\eqref{sec3:eq:gamma_long} to define the total self--energy, 
$\Sigma=\Sigma_\mathrm{long}+\Sigma_\mathrm{trans}$. This can 
be introduced in Eq.\eqref{sec3:eq:dys1} and Eq.\eqref{sec3:eq:eom_G} to obtain the final form of the equation of motions for the $G$ written as time--derivative
\begin{align}
\label{sec3:eq:eom_G_with_se}
\[\mathrm i \partial _{t_1} - h_\mathrm{ext}\(1\) \]G\(1_\ga,2_\gb\) =\delta\(1_\ga-2_\gb\) + \Sigma\(1_\ga,3_\gc\) G\(3_\gc,2_\gb\),
\end{align}
and as Dyson--like equation
\begin{align}
\label{sec3:eq:dys_with_se}
G\(1_\ga,2_\gb\) =G_0\(1_\ga,2_\gb\) + G_0\(1_\ga,3_\gc\)\Sigma\(3_\gc,4_\eta\) G\(4_\eta,2_\gb\).
\end{align}
\end{widetext}
Eq.\eqref{sec3:eq:eom_G_with_se} and Eq.\eqref{sec3:eq:dys_with_se} are equivalent and represent two different formulations of the well--known
KBE. 

From the equation of motion for the scalar and for the vector potential we can now derive the equations governing the response functions. Indeed these functions describe the change in the 
observables due to the total (external plus induced) perturbations and provide a connection between Eq.\eqref{sec2:eq:V_n},
Eq.\eqref{sec3:eq:eom_A} and Eq.\eqref{sec3:eq:eom_G}.
We analyse, now, separately the two contributions: transverse (photon--induced) and longitudinal (electronic and phonon--mediated).

\subsubsection{The photon--induced response function}
\label{sec3C1}
The photon--induced response function represents the
transverse photon propagator. We start from Eq.\eqref{sec3:eq:eom_A} and divide the total current in an external and an induced part: 
$\mathbf J_\mathrm{tot} = \mathbf J_\mathrm{ext} + \mathbf J_\mathrm{ind}$. Thus, we can express the solution of Eq.\eqref{sec3:eq:eom_A} as
\begin{align}
\braket{\hat{\mathbf{A}}\(1\)}_C = -\frac{4\pi}{c}\int \mathrm d2\, \overleftrightarrow{\mathcal D}_0\(1,2\)\[\mathbf J_\mathrm{ext}\(2\) + \mathbf J_\mathrm{ind}\(2\)\],
\end{align}
where $\mathcal D_0\(1,2\)$ is the free transverse photon propagator, solution of 
\begin{align}
\(\frac{1}{c^2}\partial^2_{t_1} - \nabla_1^2\)\overleftrightarrow{\mathcal D}_0\(1,2\) =\overleftrightarrow{\delta}^\perp\(1,2\).
\end{align}
By applying the chain rule, Eq.\eqref{sec3:eq:dgg} we derive the equation of motion for the transverse photon propagator 
\begin{align}
\label{sec2:eq:dyson_D}
\overleftrightarrow{\mathcal D}\(1,2\)=
\overleftrightarrow{\mathcal D}_0\(1,2\)+\overleftrightarrow{\mathcal D}_0\(1,3\) \overleftrightarrow{P}\(3,4\) \overleftrightarrow{\mathcal D}\(4,2\),
\end{align}
where we have defined the transverse photon polarisation $\overleftrightarrow{P}\(3,4\)$:
\begin{align}
\label{sec3:eq:trans_pol}
\overleftrightarrow{P}\(1,2\) =-\frac{4\pi}{c}\frac{\delta \mathbf J_\mathrm{ind}\(1\)}{\gd \braket{\hat{\mathbf{A}}\(2\)}_C}.
\end{align}
To obtain the equation for the transverse polarisation we need a microscopic expression for the induced current, $\mathbf J_\mathrm{ind}$. This is related to the
electronic Green's function by
\begin{multline}
\JJ_\mathrm{ind}\(1\) = \braket{\hat{\JJ}\(1\)}_C - \braket{\hat\rho\(1\)\hat{\mathbf A}\(1\)}_C \\
=\mathrm i\mathbf\Pi\(1,1'\) G\(1,1'\) |_{1'=1^+}.
\label{eq:J_ind}
\end{multline}
In deriving Eq.\eqref{eq:J_ind} we have omitted the $\rho \mathbf A$ term as we are considering averages on states with a fixed population of photons
such that $\la d^{\dagger}_{\qq+\GG,\gl}\ra_C=\la d_{\qq+\GG,\gl}\ra_C=0$. This implies that $\braket{\hat\rho\(1\)\hat{\mathbf A}\(1\)}_C=0$.
From Eq.\eqref{sec3:eq:trans_pol} and Eq.\eqref{sec2:eq:trans_prop} we finally get a closed
 expression for the transverse photon polarisation
\begin{align}
\label{sec3:eq:t_pol}
P_{\ga\gb}\(1,2\) = -\mathrm i\Pi_\ga\(1,1'\)G\(1,3\)\Gamma_\gb\(3,4,2\)G\(4,1'\)|_{1'=1}.
\end{align}

\subsubsection{The phonon--induced response function}
\label{sec3C2}
We now look into the screened Coulomb interaction $W$ defined in Eq.\eqref{sec3:eq:W}.
By using Eq.\eqref{sec3:eq:u} we obtain
\begin{widetext}
\begin{align}
\label{sec3:eq:dyson_w}
W\(1,2\) = w_0\(1,2\) +
w_0\(1,2\) \[\frac{\delta \braket{\hat{\rho}({3})}_C}{\delta\rho_\mathrm{ext}(2)} - \frac{\delta \braket{\hat{N}({3})}_C}{\delta\rho_\mathrm{ext}(2)}\] 
= w_0(1,2) + w_0(1,3)\[p_\mathrm e(3,4)W(4,2) - \frac{\delta \braket{\hat{N}({3})}_C}{\delta\rho_\mathrm{ext}(2)} \],
\end{align}
\end{widetext}
where
$w_0(1,2) = \delta(t_1-t_2)/|\mathbf r_1 - \mathbf r_2|$
 is the bare Coulomb interaction, and $p_\mathrm e$ is
the longitudinal electronic polarisation:
\begin{align}
p_\mathrm e(1,2)=\frac{\delta \braket{\hat{\rho}({1})}_C}{\delta U(2)}.
\end{align}
Eq.\eqref{sec3:eq:dyson_w} includes two contributions. One is coming from the electronic density (via $p_e$) and the other from the nuclear density (via $\gd N/\gd \gr$). Thus, 
by following Ref.\onlinecite{Leeuwen2004a}, we separate $W$ into an electronic plus a nuclear part.

The electronic polarisation is still defined, as in the purely electronic case\cite{strinati}, in terms of the electronic component of the total density 
\begin{align}
 \braket{\hat{\rho}({1})}_C = -\mathrm iG(1,1^+),
\end{align}
from which it immediately follows that
\begin{align}
\label{sec3:eq:l_vertex}
p_\mathrm e(1,2) = \mathrm i G(1,2)\gamma(3,4,2)G(4,1).
\end{align}
Again, in the derivation of Eq.\eqref{sec3:eq:l_vertex} we used Eq.\eqref{sec3:eq:dgg}.

The new contribution in Eq.\eqref{sec3:eq:dyson_w} is due to the change in the nuclear density induced by a change of the purely electronic part of 
the external charge. Indeed, this is the source of e--p interaction that describes the link between the nuclear motion 
and the electronic dynamics:
\begin{align}
\label{sec3:eq:e_n_cf}
\frac{\delta \braket{\hat{N}({3})}_C}{\delta\rho_\mathrm{ext}(2)}.
\end{align}
In order to link this quantity to the microscopic correlation functions we need to introduce the
nuclear density--density correlation function, $D$, given by
\begin{align}
D(1,2) = -\mathrm i\braket{\Delta\hat{N}( 1)\Delta\hat{N}( 2)}_C,
\end{align}
where the fluctuation of an operator is expressed as $\Delta\hat{O} = \hat{O} - \braket{\hat{O}}$. 

It is now possible to define an equation for $D$ thanks to the introduction, in Eq.\eqref{sec2:eq:h_ext}, of the external nuclear charge $N_\mathrm{ext}$. Indeed we start by noticing that, as
the interaction with the electronic density is 
$\int \mathrm d^3r\, \hat\phi(\mathbf r,t)\rho_\mathrm{ext}(\mathbf r, t)$, it follows that
\begin{align}
\label{sec3:eq:delta_Na}
\frac{\delta\braket{\hat{N}({1})}_C}{\delta\rho_\mathrm{ext}({2})} =-\mathrm i\braket{\Delta\hat{N}( 1)\Delta\hat\phi( 2)}_C.
\end{align}
If now we use the solution of the Poisson equation to rewrite $\Delta\hat\phi$ in terms of charge variations, we get
\begin{multline}
\label{sec3:eq:delta_N}
\frac{\delta\braket{\hat{N}({1})}_C}{\delta\rho_\mathrm{ext}({2})} 
= \mathrm i w_0(2,3)\[\braket{\Delta\hat{N}( 1)\Delta\hat{N}( 3)}_C -
\capo
\braket{\Delta\hat{N}( 1)\Delta\hat{\rho}(3)}_C\].
\end{multline}
Using the same procedure of above we can evaluate a similar derivative, $\frac{\delta\braket{\hat{n}(1)}}{\delta N_\mathrm{ext}( 2)}$. In this case the interaction
with $ N_\mathrm{ext}$ is $\int \mathrm d^3r\, \hat{V}_\mathrm n(\mathbf r)N_\mathrm{ext}(\mathbf r,t)$. It follows, then, that
\begin{align}
\frac{\delta\braket{\hat{n}({1})}_C}{\delta N_\mathrm{ext}({2})} =-\mathrm i\braket{\Delta\hat{n}( 1)\Delta\hat{V}_{\mathrm n}( 2)}_C.
\end{align}
By using again the solution of the Poisson equation we get that
\begin{align}
\frac{\delta\braket{\hat{N}(3)}}{\delta \rho_\mathrm{ext}(2)}=-\frac{\delta\braket{\hat{n}(3)}}{\delta N_\mathrm{ext}(2)}.
\label{sec3:eq:delta_N_equality}
\end{align}
\begin{widetext}
We can, now, easily evaluate Eq.\eqref{sec3:eq:e_n_cf} via Eq.\eqref{sec3:eq:delta_N_equality}. 
Taking into account the definition of $\hat n$ we can use the functional derivative chain rule to obtain a Dyson-like equation
\begin{multline}
w_0(1,3)\frac{\delta\braket{\hat{n}(3)}}{\delta N_\mathrm{ext}(2)} = w_0(1,3)D(3,4)w_0(4,2) + w_0(1,3)\frac{\delta\braket{\hat{\rho}(3)}}{\delta N_\mathrm{ext}(2)}= \\
=w_0(1,3)D(3,4)w_0(4,2) + w_0(1,3)\frac{\delta\braket{\hat{\rho}(3)}}{\delta U(4)} \frac{\delta U(4)}{\delta \braket{\hat{n}(5)}} 
\frac{\delta\braket{\hat{n}(5)}}{\delta N_\mathrm{ext}(2)}=\\
=w_0(1,3)D(3,4)w_0(4,2) + w_0(1,3)p_\mathrm e(3,4)w_0(4,5)\frac{\delta\braket{\hat{n}(5)}}{\delta N_{\m ext}(2)}.
\end{multline}
Iterating the equation leads to the following solution
\begin{align}
w_0(1,3)\frac{\delta\braket{\hat{n}(3)}}{\delta N_\mathrm{ext}(2)} = \[1 - w_0p_\mathrm e\]^{-1}(1,3)w_0(3,4)D(4,5)w_0(5,2).
\end{align}
which we can then replace in Eq.\eqref{sec3:eq:dyson_w} obtaining the final expression for the total screened interaction
\begin{align}
W(1,2) = w_0(1,2) + w_0(1,3)p_\mathrm e(3,4)W(4,2) + 
 \[1 - w_0p_\mathrm e\]^{-1}(1,3)w_0(3,4)D(4,5)w_0(5,2),
\end{align}
\end{widetext}
and solving it in order to $W$, which leads to two separate contributions
\begin{align}
W(1,2) = W_\mathrm e(1,2) + W_\mathrm{ph}(1,2).
\end{align}
The first term, $W_\mathrm e$, accounts for the interactions between the electrons and has no direct contribution from the nuclei
\begin{align}
W_\mathrm e(1,2) = [1-w_0p_\mathrm e]^{-1}(1,3)w_0(3,2),
\end{align}
while the second term introduces the effects of the and the contribution from the nuclear density fluctuations
\begin{align}
W_\mathrm{ph}(1,2) = [1-w_0p_\mathrm e]^{-1}(1,3)W_\mathrm e(3,4)
D(4,5)w_0(5,2).
\end{align}

\subsubsection{The vertex functions}
\label{sec3D}
In equations \eqref{sec3:eq:deltaG/R} and \eqref{sec3:eq:deltaG/J} we have introduced a longitudinal and a transverse vertex function. The equations of motion that govern their dynamics can
be easily found by differentiating the inverse of the Dyson equation, Eq. \eqref{sec3:eq:dys1}
\begin{align}
G^{-1}(1,2) = G^{-1}_0(1,2) - \Sigma(1,2).
\end{align}
Thus we can write for the longitudinal vertex, which is given by Eq.\eqref{sec3:eq:gamma_long}, the following equation
\begin{multline}
\label{sec3:eq:long_V}
\gamma(1,2,3) = -\frac{\delta G^{-1}(1,2)}{\delta U(3)}=
 \delta(1,2)\delta(1,3) +\\
\frac{\delta\Sigma(1,2)}{\delta G(4,5)}G(4,6)\gamma(6,7,3)G(7,5),
\end{multline}
and for the transverse part (Eq.\eqref{sec2:eq:trans_prop}) we have that
\begin{multline}
\label{sec3:eq:trans_V}
\mathbf\Gamma(1,2,3) = {\mathbf \Pi}(1)\delta(1,3)\delta(1,2) \\+
\frac{\delta\Sigma(1,2)}{\delta G(4,5)}G(4,6)\mathbf\Gamma(6,7,3)G(7,5),
\end{multline}
Vertex and response functions are the ingredients we need to close the equation of motion for the Green's function in a set of integro--differential equations. At this stage is important to note that
trough the functional derivative $\frac{\delta\Sigma}{\delta G}$ all interactions mix together in a complex dynamics.

\subsection{Final form of the Hedin's equations}
\label{sec3E}
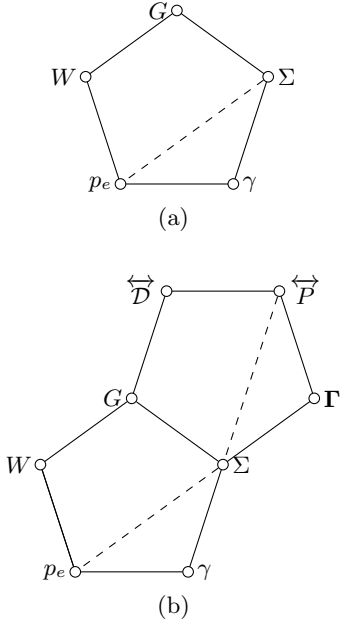
\begin{figure}
\begin{center}
 \begin{tikzpicture}
  \tikzstyle{every node}=[draw,circle,fill=white,minimum size=4pt,
              inner sep=0pt]
    \draw (0,0) node (1) [label=left:$G$] {}
    -- ++(324:1.5cm) node (2) [label=right:$\Sigma$] {}
    -- ++(252:1.5cm) node (3) [label=right:$\gamma$] {}
    -- ++(180:1.5cm) node (4) [label=left:$p_e$] {}
    -- ++(108:1.5cm) node (5) [label=left:$W$] {}
    -- (1);
    \draw[dashed] (2) -- (4);
 \end{tikzpicture}
 \\
 (a)
 \\
 \vspace{.5cm}
 \begin{tikzpicture}
  \tikzstyle{every node}=[draw,circle,fill=white,minimum size=4pt,
              inner sep=0pt]
	\draw (0,0) node (1) [label=left:$G$] {}
    -- ++(72:1.5cm) node (2) [label=left:$\overleftrightarrow{\mathcal D}$] {}
    -- ++(0:1.5cm) node (3) [label=right:$\overleftrightarrow P$] {}
    -- ++(288:1.5cm) node (4) [label=right:$\mathbf \Gamma$] {}
    -- ++(216:1.5cm) node (5) [label=right:$\Sigma$] {}
    -- ++(252:1.5cm) node (6) [label=right:$\gamma$] {}
    -- ++(180:1.5cm) node (7) [label=left:$p_e$] {}
	-- ++(108:1.5cm) node (11) [label=left:$W$] {}
    -- (1);
	\draw (1) -- (5)
		(7) -- (11);
	\draw[dashed] (3) -- (5)
		(5) -- (7);
 \end{tikzpicture}
 \\
 (b)
 \\
\end{center}
\caption{\footnotesize{Schematic representation of the different formulations of Hedin's equations.
(a) Longitudinal case. This is well known and widely used in the literature in its equilibrium and
out--of--equilibrium versions. When the interaction with the quantised electromagnetic field is switched on
a new pentagon must be added [case (b)] where the corners are the photon propagator and the transverse electronic 
polarisation and vertex function. The dashed lines correspond to the generalised $GW$ approximation where all vertex functions are neglected.
}}
\label{fig:3}
\end{figure}

We are now ready to present the full set of the Hedin's equations. Those are schematically showed in Fig.\ref{fig:3}. The first equation gives us the electron propagator $G$
\begin{align}
\label{sec3:eq:g_hedin}
G(1,2) = G_0(1,2) + G_0(1,3)\Sigma(3,4)G(4,2).
\end{align}
The second equation allows us to obtain the self-energy
\begin{widetext}
\begin{align}
\label{sec3:eq:sigma_hedin}
\Sigma(1,2) = \mathrm i\[ G(1,3)\gamma(3,2,4)W(4,1^+) +
\left. \sum\limits_{\ga,\gb=1}^3\Pi_\ga(1,1')G(1,3)\Gamma_\gb(3,2,4)\mathcal D_{\gb\ga}(4,1')\]\right|_{1'=1},
\end{align}
\end{widetext}
in which a new term on the right hand side is present, due to the fact that the electromagnetic field is now quantised. The 
fourth and fifth equations describe the screened Coulomb interaction
\begin{align}
W(1,2) = W_\mathrm e(1,2) + W_\mathrm{ph}(1,2),
\end{align}
and the transverse photon propagator 
\begin{align}
\label{sec2:eq:dyson_D_1}
\overleftrightarrow{\mathcal{D}}(1,2) =\overleftrightarrow{\mathcal D}_0(1,2) + \overleftrightarrow{\mathcal D}_0(1,3)\overleftrightarrow{P}(3,4)\overleftrightarrow{\mathcal D}(4,2).
\end{align}
In order to evaluate them we need the longitudinal and transverse polarisations
\begin{align}
p_\mathrm e(1,2) = \mathrm i G(1,2)\gamma(3,4,2)G(4,1),
\end{align}
and,
\begin{align}
P_{\ga\gb}(1,2) = \mathrm i\Pi_\ga(1,1')G(1,3)\Gamma_\gb(3,4,2)G(4,1')|_{1'=1}.
\end{align}
These equations form the sixth and seventh Hedin's equations that introduce the corresponding vertex functions, whose equations of motion represent the last two Hedin equations:
\begin{multline}
\label{sec3:eq:long_v}
\gamma(1,2,3) = \delta(1,2)\delta(1,3) \\
+\frac{\delta\Sigma(1,2)}{\delta G(4,5)}G(4,6)\gamma(6,7,3)G(7,5),
\end{multline}
and
\begin{multline}
\label{sec3:eq:trans_v}
\mathbf\Gamma(1,2,3) = {\mathbf\Pi}(1)\delta(1,3)\delta(1,2) \\+
\frac{\delta\Sigma(1,2)}{\delta G(4,5)}G(4,6)\mathbf\Gamma(6,7,3)G(7,5).
\end{multline}
The only missing term is the equation of motion for $D$, the nuclear density--density correlation function. However it is well known~\cite{Leeuwen2004a} that
the electronic and nuclear parts of Hedin's equations can be safely decoupled. As a consequence the $D$ propagator can be thought as to be given and the purely
electronic and photonic components of Hedin's equations can be solved self--consistently.


\begin{figure}[h]
\begin{center}
\epsfig{figure=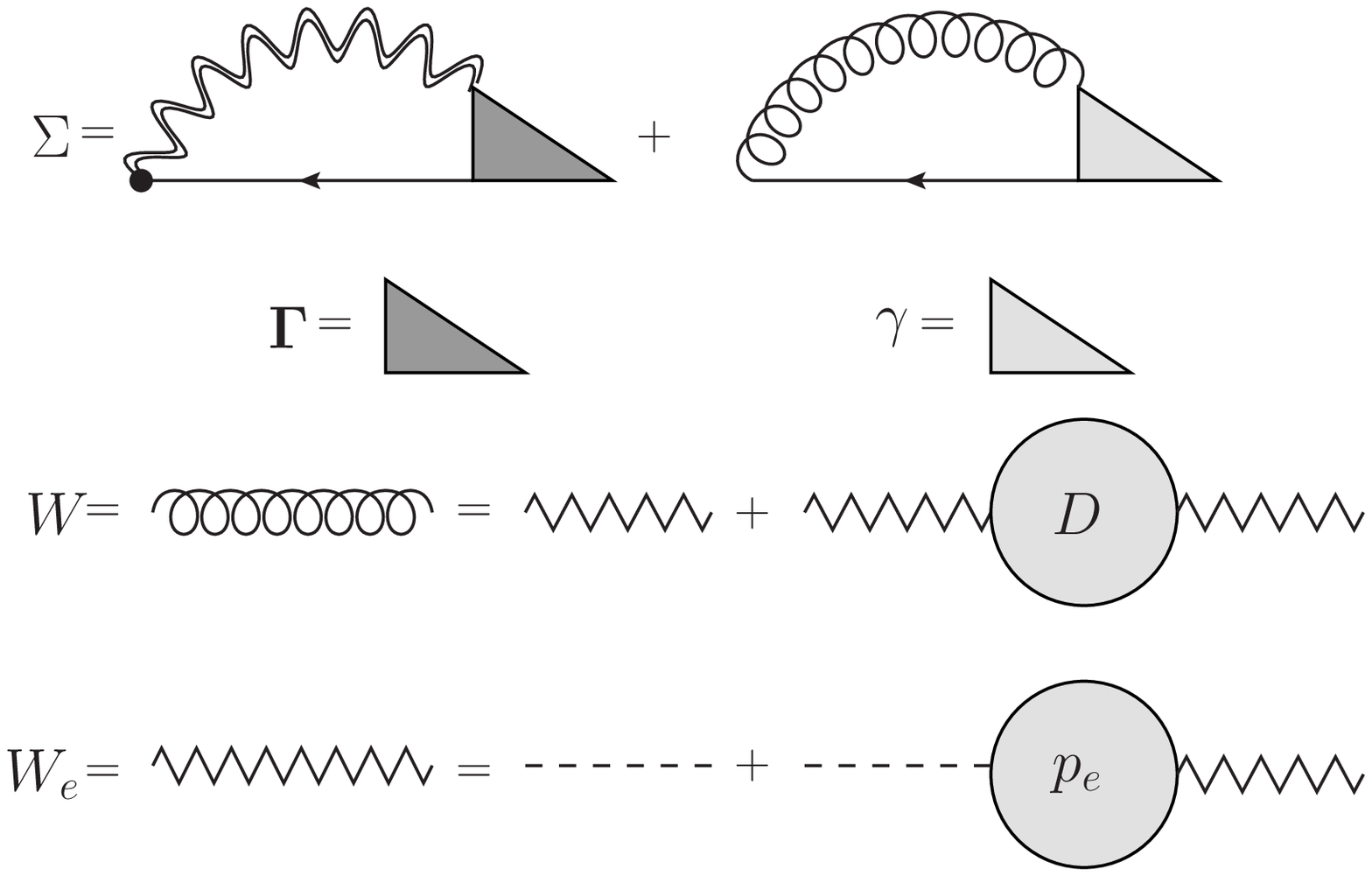,width=8cm}\\
\epsfig{figure=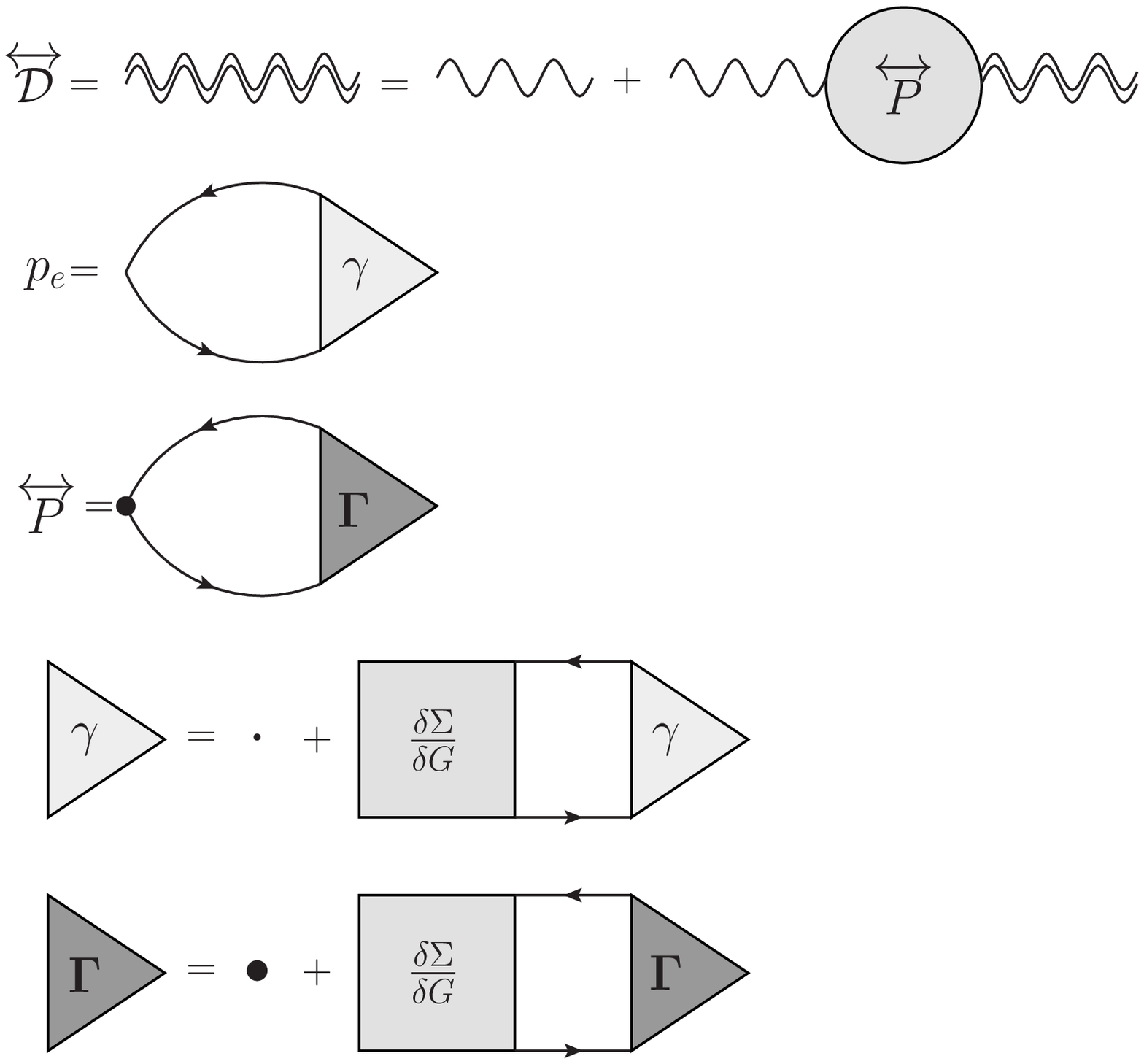,width=8cm}
\caption{\footnotesize{Diagrammatic representation of the Hedin's equations. The $\bullet$ represents a $\mathbf{\Pi}$ differential operator, defined in Eq.\eqref{sec3:eq:Pi}.}}
\label{fig:3a}
\end{center}
\end{figure}

We have represented the new set of Hedin's equations in diagrammatic form in Fig.~\ref{fig:3a} and in schematic form
in Fig.~\ref{fig:3}. This last representation also provides a method to solve the equations self-consistently. Starting from a reasonable approximation
for the longitudinal and transverse parts of the self-energy $\Sigma$, we can then evaluate the longitudinal and the transverse vertex, respectively $\gamma$
and $\mathbf \Gamma$. Following that, we can obtain the longitudinal and transverse polarisations, $p_e$ and $\overleftrightarrow P$ and incorporate the results in the
photon propagator $\overleftrightarrow{\mathcal D}$ and, assuming that we have a good description for the nuclear oscillations in the form of the phonon propagator $D$, the screened
longitudinal interaction $W$. As the last step of the cycle, we can obtain the expression for the Green's function, $G$. This is analogous to the usual
description for a self-consistent MBPT calculation, only that in this case we have two sub-cycles to evaluate: one for the transverse and another for the
longitudinal electromagnetic field.

\section{The $GW$ approximation}
\label{sec4}
The $GW$ approximation~\cite{GW_review} is based on the assumption that the corrections to the vertex can be ignored. In the present case it corresponds
to ignore the second term in Equations~\eqref{sec3:eq:long_v} and \eqref{sec3:eq:trans_v} so that the vertex functions acquire a simple form:
\begin{align}
\mathbf\Gamma^{GW}(1,2,3) = {\mathbf\Pi}(1)\delta(1,3)\delta(1,2),
\end{align}
and
\begin{align}
\gamma^{GW}(1,2,3) = \delta(1,2)\delta(1,3).
\end{align}
As a consequence both the longitudinal and transverse polarisation functions turn into an independent--particle representation
\begin{align}
\label{sec4:eq:long_p_gw}
p_\mathrm e^{GW}(1,2) = \mathrm i G(1,2)G(2,1),
\end{align}
and
\begin{align}
\label{sec4:eq:trans_p_gw}
P_{\ga\gb}^{GW}(1,2) = \mathrm i\Pi_\ga(1,1')G(1,2)\Pi_\gb(2)G(2,1')|_{1'=1}.
\end{align}
The final expression for the electronic self-energy in the $GW$ approximation is then given by
\begin{multline}
\Sigma^{GW}(1,2) = \mathrm i\[ G(1,2)W(2,1) + \right.\\
\left.\left. \sum\limits_{\ga,\gb=1}^3\Pi_\ga(1,1')G(1,2)\Pi_\gb(2)\mathcal D_{\gb\ga}(2,1')\]\right|_{1'=1}.
\label{sec4:eq:gw}
\end{multline}
The longitudinal $GW$ self--energy has been extensively studied and its formulation for the e--e
and e--p parts are well known.

\begin{figure}[h]
\begin{center}
\epsfig{figure=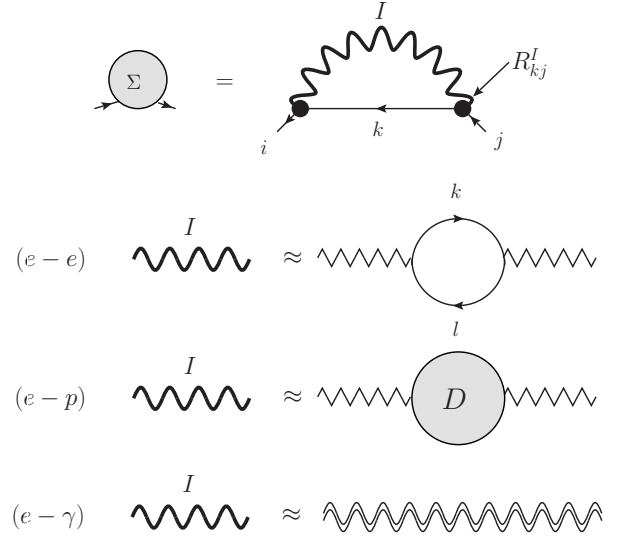,width=8cm}
\caption{\footnotesize{
The generalised $GW$ approximation for the three kind of interactions: e--e, e--p, and e--$\gc$. 
The wiggled propagator in the e--e case represents a statically screened
interaction, as explained in Ref.\onlinecite{m.2012}.
}}
\label{fig:4}
\end{center}
\end{figure}

\section{A simplified formulation using a single--time density matrix representation}
\label{sec5}
The BKE, Eq.\eqref{sec3:eq:eom_G_with_se}, are very hard to solve for practical applications and in realistic materials.
The reason is the complex two--times dependence and spatial non--locality that enormously increases the complexity of the problem compared to the
more common methods used in the \emph{ai}--MBPT scheme~\cite{Onida2002}. 

An approach that is attracting great interest is based on the reduction of the complex equation for $G\(1_\ga,2_\gb\)$ in a closed equation for the 
density matrix, $\gr\(1\)\equiv-i G\(1_-,1_+\)=-i G^{<}\(1,2\)$. This approach is based on the GBKA and, in the most
recent approaches, on the CCA.

The aim of this section is to extend the derivation of the reduced KBE to the general case of the simultaneous presence
of e--e, e--p, and e--$\gamma$ interactions.

We start by expressing all quantities in the second quantisation basis
\begin{gather}
G\(1,2\)=\sum_{i,j} \phi^{*}_i\(\rr_1\) \phi_j\(\rr_2\) G_{ij}\(t_1,t_2\),
\label{eq:cca_1}\\
\Sigma\(1,2\)=\sum_{i,j} \phi^{*}_i\(\rr_1\) \phi_j\(\rr_2\) \Sigma_{ij}\(t_1,t_2\),
\label{eq:cca_2}
\end{gather}
in such a way to remove the spatial dependence and concentrate our attention on the time arguments. We will discuss in Sec.\ref{sec8} the properties of the
single--particle basis used in \e{eq:cca_2}.

In this single--particle basis the KBE for $G^{<}\(t,t'\)$ can be rewritten in a compact, matrix--like form:
\begin{widetext}
\begin{align}
\label{eq:cca_3}
\[\mathrm i \partial _{t} - h_\mathrm{ext}\(t\) \]G^{<}\(t,t'\) = \int\,d\bar{t} \[ \Sigma^{r}\(t,\bar{t}\)G^{<}\(\bar{t},t'\)
+\Sigma^{<}\(t,\bar{t}\)G^{a}\(\bar{t},t'\)\right]+{\rm h.c.},
\end{align}
and its adjoint
\begin{align}
\label{eq:cca_3p}
\[\mathrm i \partial _{t'} - h_\mathrm{ext}\(t'\) \]G^{<}\(t,t'\) = \int\,d\bar{t} \[ G^{r}\(t,\bar{t}\)\Sigma^{<}\(\bar{t},t'\)
+G^{<}\(t,\bar{t}\)\Sigma^{a}\(\bar{t},t'\)\right]+{\rm h.c.},
\end{align}
In Eq.\eqref{eq:cca_3} and Eq.\eqref{eq:cca_3p} the retarded/advanced functions carry a superscript (r)/(a) and are 
defined in terms of the lesser and greater functions according to
\begin{align}
\label{eq:cca_ra}
X^{\rm (r)}(t,t')=[X^{\rm (a)}(t',t)]^{\dagger}=\gt(t-t')\left[X^{>}(t,t')-X^{<}(t,t')\right],
\end{align}
where $X$ can be $G$, $\Sigma$ or any other two--time correlation function. 

By adding the Eqs.~\eqref{eq:cca_3} and~\eqref{eq:cca_3p}, and taking the derivative on the macroscopic time--axis, $T=\(t+t'\)/2$ we get the equation of motion for $\gr\(T\)=-i G^{<}\(T,T\)$
\begin{align}
\frac{d}{dT}\gr\(T\)+\mathrm i\left[h_\mathrm{ext}\(T\),\gr\(T\)\right]=-S\[\{G\},\{\Sigma\}\]\(T\),
\label{eq:cca_4}
\end{align}
where all the self--energy effects are embodied in the complex collision integral, $S\[\{G\},\{\Sigma\}\]\(T\)$ that is still a function of the $\lessgtr$ two--times Green's functions and self-energies:
\begin{align}
\label{eq:cca_5}
S\[\{G\},\{\Sigma\}\]\(T\) = \int\,dt \[ \Sigma^{>}\(T,\bar{t}\)G^{<}\(\bar{t},T\) +G^{<}\(T,\bar{t}\)\Sigma^{>}\(\bar{t},T\)+\(\lessgtr \rightleftarrows \gtrless\)\].
\end{align}
In Eq.\eqref{eq:cca_5} the $\(\lessgtr \rightleftarrows \gtrless\)$ indicates that the second part of the integral is obtained from the first part 
by exchanging $>$($<$) with $<$($>$).

The collision integral includes contributions from the local and non--local self--energies $\Sigma\(t,t'\)=\Sigma_s\(t\)+\Sigma_c\(t,t'\)$. 
The local part of the self--energy defines a coherent part of $S$ moving out of the time integral: $S\[\{G\},\{\Sigma\}\]\(T\)=S ^\mathrm{coh}\(T\)+S ^\mathrm{dyn}\[\{G\},\{\Sigma\}\]\(T\)$.

The coherent part is embodied in the $h_\mathrm{ext}\(t\)$ term which turns into
\begin{align}
\mathrm i\left[h_\mathrm{ext}\(T\),\gr\(T\)\right]+S\[\{G\},\{\Sigma\}\]\(T\)=\mathrm i\left[h\(T\),\gr\(T\)\right]+S ^\mathrm{coh}\[\gr\]\(T\) +S ^\mathrm{dyn}\[\{G\},\{\Sigma\}\]\(T\),
\label{eq:cca_5p}
\end{align}
with
\begin{align}
S ^\mathrm{coh}\[\gr\]\(T\)= \mathrm i\[\Sigma_s\(T\),\gr\(T\)\].
\label{eq:cca_5s}
\end{align}
\end{widetext}
The different possible approximations to $\Sigma_{s}$
reflect the different kind of physics introduced in the dynamics and can already account for important effects. Different cases can be considered:

(i) A mean--field potential that mimics the correlation effects. An example is
DFT, where $\Sigma_{s}\(t\)$ is local in space and given by the sum of the 
Hartree and exchange--correlation potential.

(ii) HF self--energy. In this case no correlation is included. The Hartree-Fock (HF) self--energy reads:
\begin{align}
\[\Sigma_s\(t\)\]_{pq}=V_{\substack{ pq\\mn}}\gr_{nm}\(t\),
\end{align}
with the four-index tensor ${V_{\substack{ ij\\ mn}}=2\[w_0\]_{imnj}-\[w_0\]_{imjn}}$ and 
$\[w_0\]_{imnj}$ the two--electron bare Coulomb integrals:
\begin{align}
\[w_0\]_{imnj}\equiv \int d\rr\,d\rr' \phi^{*}_i\(\rr\) \phi^{*}_m\(\rr'\) w_0\(\rr-\rr'\) \phi_n\(\rr'\) \phi_j\(\rr\).
\end{align}

(iii) Hartree plus a Coulomb Hole and Screened Exchange\,(COHSEX) self--energy.
In this case correlation is included using a linear--response 
approximation but dynamical effects are neglected. The COHSEX 
self--energy reads
\begin{align}
\[\Sigma_s\(t\)\]_{pq}=V_{\substack{ pq\\mn}}\(t\)\gr_{nm}\(t\),
\label{eq:cca_5a}
\end{align}
with, now $V\(t\)_{\substack{ ij\\ mn}}=2\[w_0\]_{imnj}-W\[\gr\(t\)\]_{imjn}$ and $W$ the screened Coulomb hole
potential. This is routinely calculated in the random--phase approximation\,(RPA) where
\begin{align}
W\[\gr\]\(\rr,\rr'\) \equiv \int\,d\overline{\rr}\, \gee^{-1}_{\rm RPA}\[\gr\]\(\rr,\overline{\rr}\) w_0\(\overline{\rr}-\rr'\).
\end{align}

The choice of the local part of the self--energy is essential as it already describes a large part of the level of correlation 
embodied in the many--body dynamics. It has been formally proved, for example, that the COHSEX self--energy
describes in the linear regime the excitonic effects and reduces Eq.\eqref{eq:cca_4} to the well known
Bethe--Salpeter equation~\cite{Attaccalite2011}.

The collision integral $S ^\mathrm{dyn}\(t\)$ is
non--local in time and its 
functional form is uniquely determined once an approximation for the 
correlation self--energy, $\Sigma_{c}$, is made.

Let us consider here the $GW$ approximation, Eq.\eqref{sec4:eq:gw}, that can be rewritten, by using Eq.\eqref{eq:cca_1}, in a very general and compact form as
\begin{align}
\Sigma^{\lessgtr}_{ij}\(t,t'\) = \mathrm i \sum_I \[\(R^I\)^{\dag} G^{\lessgtr}\(t,t'\) R^I\]_{ij} W^{\lessgtr}_{I}\(t,t'\).
\label{eq:cca_6}
\end{align}
In Eq.\eqref{eq:cca_6} $R^I$ and $G$ are matrices [see Eqs.~\e{eq:cca_1} and~\e{eq:cca_2}] and the product in the square parenthesis represents a matrix multiplication. 
In Eq.\e{eq:cca_6}, and from this point onward, we use the Einstein convention that repeated indices are summed over. 
$W_I$, instead, is a scalar function. $R_I$ and $W$ have different definitions depending on the kind of interaction they are describing. 

(a) In the e--p case $I=\(\qq\gl\)$ and represents
the phonon branch\,($\gl$) and momentum\,($\qq$) pair. It follows that
\begin{align}
R^I_{ij}=R^{\(\qq\gl\)}_{ij}\equiv \int\,d\rr \phi_i^*\(\rr\) \partial_{\qq\gl}V_n\(\rr\) \phi_j\(\rr\),
\label{eq:cca_7}
\end{align}
with $ \partial_{\qq\gl}$ the derivative of the bare ionic potential ($V_n$) along the phonon state $\(\qq\gl\)$. Note that, if the present formalism is applied on top of DFT, 
$V_n\(\rr\)$ is replaced by the dressed self--consistent derivative of the DFT ionic potential as discussed in Ref.\onlinecite{PhysRevB.91.224310}.

(b) In the e--$\gamma$ case the $I$ index represents the photon polarisation index ($\gl$) and momentum ($\qq$) and
\begin{align}
R^I_{ij} = \int d\rr\, e^{\mathrm i\qq_I\cdot\rr} \phi^*_{i}\(\rr\) \({\bf e}_\gl \cdot \nabla\) \phi_{j}\(\rr\).
\end{align}

(c) In the e--e case the derivation of the analytic expression for $R^I$ is more mathematically involved. At the same time, however, this procedure has been 
extensively studied in the literature. We take as reference Ref.\onlinecite{Haug1992a} and Ref.\onlinecite{m.2012} to rewrite the out--of--equilibrium screened e--e interaction
as an effective interaction with a time--dependent plasma of electron--hole pairs with scattering amplitudes given by
\begin{align}
R^I_{ij} = \int d\rr\, \phi^*_{i}\(\rr\) W\[\gr\]\(\rr,\rr'\) \phi_{j}\(\rr\) \[\Phi_I\(\rr'\)\],
\end{align}
where $I=\(k,l\)$ and $\Phi_I\(\rr\)\equiv \phi^*_{k}\(\rr\) \phi_{l}\(\rr\)$. In this case, indeed, the elemental excitations are electron--hole pairs $\(k,l\)$.

In all of the three cases above, $W$ has the form
\begin{align}
W^{\lessgtr}_I\(t,t'\)\equiv W^{\lessgtr}_I\(t-t'\)\equiv \(-\mathrm i\) N_I^{\pm} e^{\pm \mathrm i\go_I \(t-t'\)}.
\label{eq:cca_8}
\end{align}
In Eq.\eqref{eq:cca_8} $N_I^{\pm}$ acquires different meanings depending on the interaction. In the e--p case it represents the occupation of the phonon bath at
a given temperature that, for simplicity, we assume to be fixed. An extended derivation can be carried on by taking into account the simultaneous set of
equations of motion for the electrons, phonons and photons. However here we would like to outline the method so we use as test case the electronic motion only. 

In the e--p and e--$\gc$ cases, then,
$N_I^{+}=N_I\(\gb\)+1$ and $N_I^{-}=N_I\(\gb\)$ with $N_I\(\gb\)$ that Bose occupation and $\gb=1/{k_B T_{el}}$ with $k_B$ the Boltzmann constant and $T_{el}$ the lattice temperature.
In the e--e case $N_I^{+}=\gr_{kk}\(t\) \(1-\gr_{ll}\(t\)\)$ and $N_I^{-}=\gr_{ll}\(t\) \(1-\gr_{kk}\(t\)\)$.

By using Eq.\e{eq:cca_8} into Eq.\e{eq:cca_6} we get that
\begin{align}
\Sigma^{\lessgtr}_{ij}\(t,t'\) = \mathrm i \sum_I \[\(R^I\)^{\dag} G^{\lessgtr}\(t,t'\) R^I\]_{ij} W^{\lessgtr}_{I}\(t-t'\).
\label{eq:cca_9}
\end{align}
\begin{figure}[h]
\begin{center}
\epsfig{figure=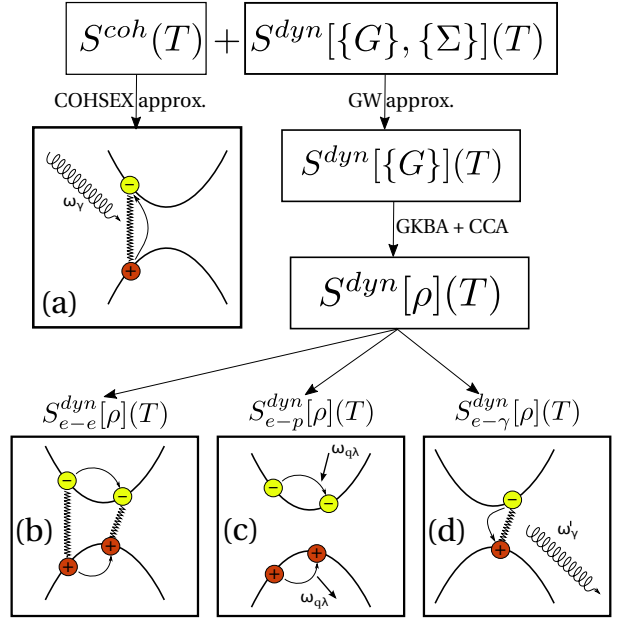,width=8cm}
\caption{\footnotesize{Schematic representation of the different links between the terms of Eq.\e{eq:exp_2} and the most elemental physical processes occurring in a typical pump--and--probe experiment:
the photo excitation\,(a), the relaxation via e--e scattering\,(b), the relaxation and dissipation via e--p scattering\,(c) and the final, slow, radiative recombination\,(d). It is crucial to note that the use
of $\Sigma_{COHSEX}$ allows to include excitonic effects (caused by the electron--hole attraction) in all the processes.
}}
\label{fig:5}
\end{center}
\end{figure}

\subsection{The Generalized Baym--Kadanoff ansatz}
\label{sec5B}
Now that $W^{\lessgtr}$ is a function of the time difference, in order to close Eq.\e{eq:cca_4} in the space of the single--time density matrices we introduce the Generalised Baym--Kadanoff Ansatz.

The GBKA
is an ansatz for $G^{\lessgtr}$ which turns it, and hence 
the collision integral, into a 
functional of $\gr$ and $G^{\rm (r)/(a)}$:
\begin{gather}
G^{<}(t,t')=-G^{\rm (r)}(t-t')\gr(t')+\gr(t)G^{\rm (a)}(t-t')       \label{eq:gkba_g}, \\
G^{>}(t,t')=+G^{\rm (r)}(t-t')\bar{\gr}(t')-\bar{\gr}(t)G^{\rm (a)}(t-t'),
\tag{\ref{eq:gkba_g}$'$}
\end{gather}
where $\bar{\gr}=1-\gr$. To transform $S ^\mathrm{dyn}\(t\)$ into a functional 
 of the density matrix, and hence to close Eq.\eqref{eq:cca_4}, one needs 
to express the propagator $G^{(\rm r)}$ in terms of $\gr$. Depending on the system there are optimal 
approximations to the propagator, the most common one being the quasi--particle\,(QP)
propagator
\begin{align}
G^{\rm (r)}(t,t')=-\mathrm i\gt(t-t')Te^{-\mathrm i\int_{t'}^{t}d\bar{t}\,h^{\rm qp}(\bar{t})}.
\label{gret}
\end{align}
For (small) finite systems the choice ${h^{\rm qp}=h^{\rm eq}}$ (usually $h^{\rm eq}$ is the HF single--particle Hamiltonian) is a good choice. For extended
systems however, the lack of damping in $h^{\rm eq}$ prevents the system to relax.  In these cases the propagator is typically corrected by adding
non--hermitian terms given by the quasi--particle life--times ${h^{\rm qp}=h^{\rm eq}+\mathrm
i\gc}$\cite{Marini2008,Haug1992,springerlink:10.1007/s100510050770,m.2012,LPUvLS.2014}.

By using Eq.\e{eq:cca_9} we rewrite $S ^\mathrm{dyn}\(T\)$ as:
\begin{widetext}
\begin{multline}
S ^\mathrm{dyn}_{ij}\[\{G\},\{\Sigma\}\]\(T\) \equiv S ^\mathrm{dyn}_{ij}\[\{G\}\]\(T\) = \mathrm i \sum_I \int\,dt \Bigl\{ \[ \(R^I\)^{\dag} G^{>}(T,t') R^I G^{<}(t',T)\]_{ij} W_I^{>}\(T-t'\)\\
+\[ G^{<}(T,t') \(R^I\)^{\dag} G^{>}(t',T) R^I \]_{ij} W_I^{>}\(t'-T\) - \(\lessgtr \rightleftarrows \gtrless\)\Bigr\},
\label{eq:cca_10}
\end{multline}
and finally, we use Eq.\e{eq:gkba_g} and Eq.(\ref{eq:gkba_g}$'$), to rewrite the first terms of Eq.\e{eq:cca_10}:
\begin{multline}
\[ \(R^I\)^{\dag} G^{>}(T,t') R^I G^{<}(t',T)\]_{ij} W_I^{>}\(T-t'\)= \\
\(-\mathrm i\)\sum_{\pm}\[ \(R^I\)^{\dag} G^{(r)}(T-t') \o{\gr}\(t'\) R^I \gr\(t'\) G^{(a)}(t'-T)\]_{ij} N_I^{\pm} e^{\pm \mathrm \go_I \(T-t'\)},
\label{eq:cca_11}
\end{multline}
with $\o{\gr}_{ij}\(T\)\equiv \gd_{ij}-\gr_{ij}\(T\)$. We now notice that, because of the time ordering of $G^{(r/a)}$, the integral in Eq.\e{eq:cca_10} runs
between $-\infty$ and $T$. 

In order to proceed with the derivation and obtain a final expression that can be easily compared with the well--known Boltzmann equation we use a drastic but
simple approximation for $G^{(r/a)}$ based on the non--interacting approximation:
\begin{align}
G^{(r)}_{ij}\(T\)\approx -\mathrm i \gt\(T\) e^{-\mathrm i\gee_i T}\gd_{ij}.
\label{eq:cca_12}
\end{align}
Eq.~\e{eq:cca_12} allows us to carry on to the final steps of the derivation in a more comfortable way by introducing the scattering matrix $S^{Is}_{ij}$ as
\begin{align}
S^{Is}_{ij}\(T\)\equiv -\mathrm i \int_{-\infty}^{T} dt \[e^{-\mathrm is\go_I t} \(R^I_{ki}\)^{*} e^{\mathrm i\(\gee_j-\gee_k\) \(T-t\)}\]
\underbrace{
\[ \o{\gr}_{kl}\(t\) R^I_{ln} \gr_{nj}\(t\) N_I^{-s}-
 \gr_{kl}\(t\) R^I_{ln} \o{\gr}_{nj}\(t\) N_I^{s}\]
}_\text{$\xi^{Is}_{kj}\(t\)$}. 
\label{eq:cca_13}
\end{align}
\end{widetext}
In Eq.\e{eq:cca_13} we have also defined the $\xi^{Is}_{kj}\(t\)$ function. The $S ^\mathrm{dyn}\(T\)$ can be finally rewritten in terms of $S$ as 
\begin{align}
S ^\mathrm{dyn}_{ij}\(T\)\equiv \mathrm i \sum_{I,\pm} \[ e^{\mathrm is\go_I t} S^{Is}_{ij}\(T\) +h.c.\].
\label{eq:cca_14}
\end{align}

\subsection{The Completed Collision Approximation}
\label{sec5C}
The time integral in Eq.\e{eq:cca_13} can be removed analytically by using the well--known Completed Collision Approximation. This is based on the {\em adiabatic} ansatz introduced
in Ref.~\onlinecite{Perfetto2015} which is based on the assumption that the characteristic time scales of the dynamics are much long compared to the time window where the 
physical properties are calculated. We will discuss more in detail the adiabatic ansatz in the next sections. Here we just formulate it by approximating
$\xi^{Is}_{kj}\(t\)\approx \xi^{Is}_{kj}\(T\)$ in Eq.\e{eq:cca_13}, so to take it outside the time integral. At this point, this can be solved analytically leading to the 
final expression for the $S$ function:
\begin{align}
S^{Is}_{ij}\(T\)\approx \(-\mathrm i\) \(R^I_{ki}\)^* \frac{e^{-\mathrm is\go_I T}}{\gee_k-\gee_j-s\go_I+\mathrm i0^+} \xi^{Is}_{kj}\(T\).
\label{eq:cca_15}
\end{align}
By plugging Eq.\e{eq:cca_15} into Eq.\e{eq:cca_14} we get the final, explicit, form of $S ^\mathrm{dyn}\(T\)$:
\begin{widetext}
\begin{align}
S ^\mathrm{dyn}_{ij}\(T\)=-\mathrm i\sum_{Is} \[ \frac{ \(R^I_{ki}\)^{*} \o{\gr}_{kl}\(T\) R^I_{ln} \gr_{nj}\(T\) N_I^{-s}-
 \(R^I\)^{\dagger}_{ik}\gr_{kl}\(T\) R^I_{ln} \o{\gr}_{nj}\(T\) N_I^{s}} {\(\gee_k-\gee_j-s\go_I-\mathrm i0^+\)} -h.c.\].
\label{eq:cca_16}
\end{align}
Eq.\e{eq:cca_16} represents an important result of this work. It shows that for any kind of interaction (e--e, e--p, e--$\gc$) the scattering term of the equation of motion for the density matrix can be rewritten in a closed
form. 
\begin{figure}[h]
\begin{center}
\epsfig{figure=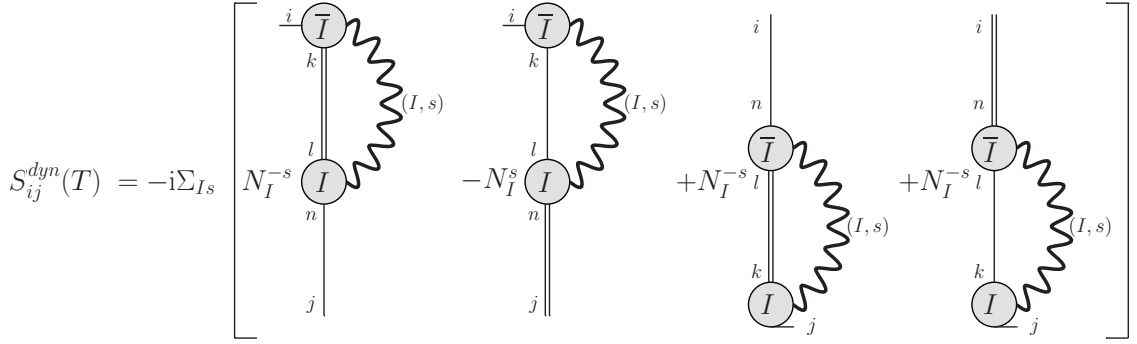,width=15cm}
\caption{\footnotesize{Graphical representation of Eq.\e{eq:cca_16}. A single line represents the density matrix $\gr$, while the double line
is its adjoint, $\o{\gr}$. The wiggled line, in the spirit of Fig.\ref{fig:4}, represents a generic interaction propagator.}}
\label{fig:6}
\end{center}
\end{figure}
\end{widetext}
Now, before moving to the next section where we will turn Eq.\eqref{eq:cca_16} into a working scheme to calculate several quantities, we want to analyse in detail
the structure of $S ^\mathrm{dyn}$ in order to draw its very general properties. Indeed, in simple terms Eq.\e{eq:cca_16} represents an elemental scattering event,
where an initial state $\(i\)$ is scattered in a series of states [$\(k\)$, $\(l\)$, $\(n\)$ and finally $\(j\)$] via the emission/absorption of an elemental
excitation (photon, e--h pair, phonon). 

In order to see this graphically we proceed as follows. We associate a density matrix $\gr_{ij}\(T\)$ with a single line connecting the state $j$ to the state
$i$, while $\o{\gr}_{ij}\(T\)$ is represented by a double line. A large circle ($\bigcirc$) will represent a generic interaction
matrix $R^I$. A label $\o{I}$ will denote an adjoint matrix $\(R^I\)^{\dagger}$. A wiggled line, instead, represents the propagation of an excitation $(I,s)$.

By using these simple rules the generic form of $S ^\mathrm{dyn}_{ij}\(T\)$ is showed in Fig.\ref{fig:6}.

Now the next step is to link the properties of $S ^\mathrm{dyn}\(T\)$ to the actual quantities that are measured in a typical pump--and--probe experiment.

The procedure that reduces the complex dynamical equation to the simple Eq.\e{eq:cca_16} form is schematically represented in Fig.\ref{fig:5}.

\section{Carrier dynamics, transient absorption and light emission in the adiabatic regime}
\label{sec6}
In the typical experiment sketched in Fig.~\ref{fig:1} a strong pump laser field is followed by a second weaker probe impulse whose 
physical properties are measured as a function of the pump--probe temporal delay. The low intensity of the probe and pump field and their temporal duration and
delay can be used to simplify further the equations. Those are the physical basis for the Low Intensity Approximation and for the introduction of the adiabatic ansatz
regime.

These two special regimes are motivated by well defined physical arguments that are introduced in this section.

The aim of this section is to use the fact that Eq.\eqref{eq:cca_16} is closed in the space of the density matrices and rewrite all possible observables relevant to the
dynamics schematically represented in Fig.\ref{fig:1} as functions of the time--dependent density matrix. 

\subsection{The adiabatic ansatz}
\label{sec6A}
The adiabatic ansatz has been first introduced in Ref.\onlinecite{Perfetto2015} and in the following we present a short
review in order to introduce the physical basis for the low--intensity and completed collision approximations. 
The temporal geometry of the pump and probe fields must appear explicitly in Eq.\e{sec2:eq:h_ext} connected
to two components of the external charge
\begin{align}
\gr_\mathrm{ext}\(1\)=\gr^{p}_\mathrm{ext}\(1\)+\gr^{P}_\mathrm{ext}\(1\),
\label{eq:exp_1}
\end{align}
with $\gr^{P/p}$ referring to, respectively, the Pump/probe components of the external charge. These terms define the Pump and Probe fields, $\EE^{P/p}\(1\)=-{\bf \nabla}
\phi_\mathrm{ext}^{P/p}\(1\)$. In a P\&p experiment we have that, in general, $|\EE^{P}|\gg |\EE^{p}|$ and the two fields are separated by a temporal delay
$\tau$. 

The delay time $\tau$ is a crucial ingredient of the dynamics. Indeed the pump field excitation induces several processes with different time--scales.
The laser excitation induces off--diagonal matrix elements of the density matrix $\gr_{ij}$, which in turn corresponds to a laser--induced
current $\JJ\(1\)=\sum_{ij} \jj_{ij}\(\rr_1\) \gr_{ji}\(t_1\)$,
with $\jj_{ij}$ the matrix elements of Eq.\e{sec2:eq:j}. The time--scale which describes the decay of the polarisation and of the current, 
$\tau_\mathrm{pol}$, is dictated by relaxation processes.

At the same time, the pump excites electrons from the valence to the conduction bands and these carriers first relax towards the band minimum\,(in the case of
electrons) and band maximum\,(in the case of holes). This relaxation occurs on a time scale $\tau_\mathrm{carr}$.

Thus, after a time $\tau_\mathrm{max}=\mathrm{max}\(\tau_\mathrm{pol},\tau_\mathrm{carr}\)$,
we may say that the system is in a quasi--stationary state. In this regime the time to relax
back to the ground state is dictated by e--$\gc$ scattering and can be of the order of picoseconds.

The key assumption of the adiabatic ansatz is that the delay $\tau$ is choosen in such a way that, for times $t\approx \tau$, we have that 
$\gr\(t + \Delta t\) \approx \gr(t)$, if $\Delta t \ll \tau_\mathrm{max}$.
The fundamental idea of such approximation is that, during the measurement process, the non--equilibrium configuration of the system is frozen. 

By using Eq.\eqref{eq:exp_1} in the external interaction term of $h$ we can now
make explicit the different terms contributing to Eq.\e{eq:cca_4}. 

\begin{widetext}
We start by using for the coherent part the COHSEX approximation, and by splitting $S ^\mathrm{dyn}$ in the three terms induced by the different possible interactions 
\begin{align}
\frac{d}{dT}\gr\(T\)+\mathrm i\left[h_\mathrm{ext}\(T\),\gr\(T\)\right]=-S ^\mathrm{coh}\[\gr\]\(T\)-S_{e-e} ^\mathrm{dyn}\[\gr\]\(T\)-S_{e-p} ^\mathrm{dyn}\[\gr\]\(T\)-S_{e-\gc} ^\mathrm{dyn}\[\gr\]\(T\).
\label{eq:exp_2}
\end{align}
\end{widetext}
The physical contents of the above equations are made clear by Eq.\e{eq:cca_16}. In all the three cases the elemental process described by $S ^\mathrm{dyn}$ is a single transition from the electronic state $i$ to 
the electronic state $j$ by spanning all possible intermediate stated composed by an electron in the level $k$ and a plasma (e--e case), phonon (e--p case) and photon (e--$\gc$) excitation. 
The fact that it is a single transition is a consequence of the use of a $GW$ approximation.

We start by noticing that, when the external Pump and probe fields are zero,
$\gr$ is diagonal:
\begin{align}
\left.\gr_{ij}\(T\)\right|_{\EE^P=\EE^p={\bf 0}}= \gd_{ij} f_i.
\label{eq:exp_3}
\end{align}
Eq.\eqref{eq:exp_2} is a non--linear equation whose non--linearity is driven by the Pump field. This non--linearity mixes the diagonal and off--diagonal components of $\gr$ 
creating a complex interplay between the induced carrier occupations and the related polarisation. However, in the case of a low--intense pumping field, if the system has reached the quasi--stationary state, we can approximate 
in the r.h.s. of Eq.\eqref{eq:cca_16}:
\begin{gather}
\gr_{kl}\(T\)\approx \gd_{kl} f_k\(T\),
\label{eq:exp_5}\\
\o{\gr}_{kl}\(T\)\approx \gd_{kl} \(1-f_k\(T\)\).
\tag{\ref{eq:exp_5}'}
\end{gather}
We will refer to this set of approximations as Low--Intensity Approximation\,(LIA). This is a well established physical regime used in a wealth of experimental setups
where the density of carriers created in the conduction bands is low enough to not alter substantially the physical properties of the material. 

\subsection{Carrier dynamics}
\label{sec6B}
In the LIA the complex structure of the dynamical kernel can be further reduced to a compact and simple form:
\begin{widetext}
\begin{align}
S ^\mathrm{dyn}_{ij}\(T\)\approx
\left. S ^\mathrm{dyn}_{ij}\(T\)\right|_\mathrm{LIA}= \mathrm i\sum_{n} \[ 
\gc^{\(-\)}_{ijn}\(T\) \gr_{nj}\(T\)+\gr_{in}\(T\) \gc^{\(+\)}_{nij}\(T\)-
\tilde{\gc}^{\(-\)}_{ijn}\(T\) \o{\gr}_{nj}\(T\)-\o{\gr}_{in}\(T\) \tilde{\gc}^{\(+\)}_{nij}\(T\)\],
\label{eq:exp_6}
\end{align}
\end{widetext}
with
\begin{gather}
\gc^{\(\pm\)}_{ijn}\(T\)=\(-\mathrm i\) \sum_{Is} \frac{ \(R^I_{ki}\)^{*} f_{k}\(T\) R^I_{kn} N_I^{-s}} {\(\gee_k-\gee_j-s\go_I\pm \mathrm i0^+\)},
\label{eq:exp_7}\\
\tilde{\gc}^{\(\pm\)}_{ijn}\(T\)=\(-\mathrm i\) \sum_{Is}\frac{ \(R^I_{ki}\)^{*} \(1-f_{k}\(T\)\) R^I_{kn} N_I^{s}} {\(\gee_k-\gee_j-s\go_I\pm \mathrm i0^+\)}.
\tag{\ref{eq:exp_7}$'$}
\end{gather}
In obtaining Eq.\e{eq:exp_6} and Eqs.\e{eq:exp_7} we have used the LIA only on the internal density matrices as their indexes are free and do not impose the
$S ^\mathrm{dyn}$ to be diagonal.

We can now link Eq.\e{eq:exp_2} to the complex dynamics we were aiming at describing at the beginning of this work, schematically represented in Fig.\ref{fig:1}. 

Physically the different contributions to $S ^\mathrm{dyn}$ represent the different channels that concur to the dynamics following the primary pump excitation. To see
in practice their effect we move to the splitting of Eq.\e{eq:exp_2} in carrier and polarisation dynamics.

The equation of motion for the carrier occupations is readily obtained by taking the diagonal components of the $\gr$, solution of Eq.\e{eq:exp_2}. In this case the
diagonal components of $S ^\mathrm{dyn}$ acquire a simple interpretation. Indeed, from Eqs.(\ref{eq:exp_6},\ref{eq:exp_7},\ref{eq:exp_7}$'$) it immediately follows that
\begin{widetext}
\begin{multline}
\left. S ^\mathrm{dyn}_{ii}\(T\)\right|_\mathrm{LIA}=
2\pi \sum_{Is k} |R^I_{ki}|^2 \[\(1-f_k\(T\)\) f_i\(T\) \gd\(\gee_i-\gee_k-s\go_I\)N_I^{-s}-\capo
f_k\(T\) \(1-f_i\(T\)\) \gd\(\gee_i-\gee_k-s\go_I\) N_I^{s}\]=\sum_I \gc_i^{\(I,e\)}\(T\)\gr_{ii}\(T\)-\gc_i^{\(I,h\)}\(T\)\o{\gr}_{ii}\(T\).
\label{eq:cd_1}
\end{multline}
\end{widetext}
Eq.~\e{eq:cd_1} represents a generic Markovian scattering of the electron/hole (labelled by the $(e/h)$ superscripts) in the state $i$ to the generic state
$k$ mediated by the emission ($s=+1$) or absorption ($s=-1$) of a generic boson of energy $\go_I$. In the e--e case this boson is an additional electron--hole
pair. In the e--p case the boson is a phonon, and in the e--$\gc$ channel it is a photon. 

In the e--p and e--e case Eq.\e{eq:cd_1} reduces to the equation derived previously by one of us~\cite{m.2012} and applied to the interpretation of the
time--resolved two--photon photon--emission experiment of bulk Silicon~\cite{Sangalli2015}. However the present case extends the derivation
to the e--p channel. This extension defines in a pure NEQ framework the radiative electron/hole lifetimes. At difference with the
usual formulation~\cite{doi:10.1021/nl503799t} the $\gc_i$ lifetimes are time--dependent and depend on the time fluctuations of the carrier
occupations. Moreover, by means of the presence of $S ^\mathrm{coh}$, the coupling with the external laser field is correctly described.

\subsection{Light absorption}
\label{sec6C}
Starting from the equilibrium condition an external field will induce an electronic dipole defined as the expectation 
value of the dipole operator, defined in terms of 
the density matrix
\begin{align}
\hat{d}={\bf \eta}_{p}\cdot {\hat{\dd}\equiv \int d\rr\; ({\bf \eta}_{p}\cdot \rr)\hat{\gr}(\rr,\rr)}=
d_{ij} \hat{\gr}_{ji},
\label{eq:exp_2a}
\end{align}
with $d_{ij}=\int d\rr\,
\varphi^*_{i}(\rr)({\bf \eta}_{p}\cdot \rr) \varphi_{j}(\rr)$ the dipole matrix elements.
For simplicity, in Eq.\e{eq:exp_2a}, we have assumed that the Pump and probe fields are polarised along the ${\bf \eta_p}$ direction.
The time--dependent expectation value of the dipole operator, $d\(T\)$, is then given by
\begin{align}
d\(T\)=\langle\Psi\T|\hat{d}|\Psi\T\rangle=d_{ij}\langle\Psi\T| \hat{\gr}_{ji}|\Psi\T\rangle,
\label{eq:exp_2b}
\end{align}
and can be calculated using the electronic polarisation function $p_\mathrm e(1,2)$ as
\begin{multline}
\gd d\T=d_{ij} \int\,dt' p_{\substack{ ji \\ lk}, \mathrm e}\(T,t'\) d_{kl} E_p\(t'\)=\\
\int dt' \[ d \circ \chi\(T,t'\) \circ d \] E_p\(t'\).
\label{eq:exp_4}
\end{multline}
In Eq.\e{eq:exp_4} we used the conventions listed in App.\ref{appA} [see Eq.\e{eq:appA_2}]. 

The absorption coefficient $\mathfrak{S}^{\tau}\(\go\)$ can be easily calculated from
the $\chi$ as\cite{Perfetto2015}
\begin{align}
\mathfrak{S}^{\tau}\(\go\)= -2\go|e\(\go\)|^{2}\Im\[d\circ \chi^{\tau}\(\go\)\circ d \],
\label{eq:trans_spectrum}
\end{align}
with $\chi^{\tau}\(\go\)$ defined in Sec.\ref{sec7} and obtained from $\chi\(t,t'\)$ by applying the Adiabatic Ansatz. The problem is now how to calculate $\chi$.

From Eq.\e{eq:exp_4} and the definition of the total (external plus induced) scalar field, Eq.\e{sec3:eq:u}, we know that the matrix components of the reducible
electronic polarisation function can be written as
\begin{multline}
\chi_{\substack{ ji\\ lk}}\(t,t'\)=
-\mathrm i\gt\(t-t'\)\\
\times \langle\Psi_{g}|
\left[\hat{c}^{\dagger}_{iH}\(t\)\hat{c}_{jH}\(t\),
\hat{c}^{\dagger}_{kH}\(t'\)\hat{c}_{lH}\(t'\)\right]|\Psi_{g}\rangle
\label{eq:abs_1},
\end{multline}
which, together with its irreducible counter--part $\tilde{\chi}$, can be rewritten as derivatives of the time--dependent density--matrix
\begin{gather}
\chi_{\substack{ ji\\ lk}}\(t,t'\)=\frac{\gd\gr_{ij}\(t\)} {\gd U_{kl}\(t'\)},
\label{eq:abs_2}\\
\tilde{\chi}_{\substack{ ji\\ lk}}\(t,t'\)=\frac{\gd\gr_{ij}\(t\)} {\gd \phi^\mathrm{ext}_{kl}\(t'\)}.
\tag{\ref{eq:abs_2}$'$}
\end{gather}
The derivation of the equation of motion for both $\chi$ and $\tilde{\chi}$ is obtained by applying the functional derivatives to the equation of motion for $\gr\T$,
\begin{widetext}
\begin{align}
\frac{d}{dt}\Bigl\{\frac{\delta\gr_{ji}\(t\)}{\delta \phi^\mathrm{ext}_{kl}\(t'\)}\Bigr\}=\frac{\delta}{\delta\phi^\mathrm{ext}_{kl}\(t'\)} \Bigl\{ 
-\mathrm i\left[h_\mathrm{ext}\(t\),\gr\(t\)\right]_{ji}-S_{ji} ^\mathrm{coh}\[\gr\]\(t\)-S_{ji} ^\mathrm{dyn}\[\gr\]\(t\)\Bigr\}.
\label{eq:abs_3}
\end{align}
The first term on the r.h.s of Eq.\e{eq:abs_3} is 
\begin{align}
\frac{\delta}{\delta\phi^\mathrm{ext}_{kl}\(t'\)} \Bigl\{ \left[h_\mathrm{ext}\(t\),\gr\(t\)\right]_{ji}\Bigr\}=\[\gd_{jk}\gr_{li}\(t\)-\gr_{jk}\(t\)\gd_{li}\]\gd\(t-t'\)+
\[h_\mathrm{ext}\(t\),\tilde\chi\(t,t'\)\]_{ \substack{ ji\\ lk} },
\label{eq:abs_4}
\end{align}
where we have used the compact form given by Eq.\e{eq:appA_4} to write the last term of Eq.\e{eq:abs_4}. The derivative of $S ^\mathrm{coh}$ can be evaluated within the COHSEX approximation
for $\Sigma_s$, Eq.\e{eq:cca_5a},
\begin{align}
\frac{\delta S ^\mathrm{coh}_{ji}\(t\)}{\delta\phi^\mathrm{ext}_{kl}\(t'\)} = \mathrm i \[\frac{\delta \Sigma^s(t)}{\delta\phi^\mathrm{ext}_{kl}\(t'\)},\gr(t)\]_{ji} =\mathrm i\[K_s\circ\tilde\chi(t,t'),\gr(t)\]_{\substack{ij\\kl}}
\label{eq:abs_5}
\end{align}
In order to evaluate the functional derivative acting on the $S ^\mathrm{dyn}$ functions we use the following chain rule: 
\begin{align}
\frac{\delta}{\delta \phi^\mathrm{ext}_{kl}\(t'\)}= \int\,d\o{t} \,\frac{\delta\gr_{mn}\(\o{t}\)}{\delta \phi^\mathrm{ext}_{kl}\(t'\)}
\frac{\delta}{\delta\gr_{mn}\(\o{t}\)},
\end{align}
so that
\begin{align}
\frac{\delta}{\delta\phi^\mathrm{ext}_{kl}\(t'\)} S ^\mathrm{dyn}_{ji}\[\gr\]\(t\)=\int\,d\o{t}\,\[K ^\mathrm{dyn}\(t,\o{t}\)\circ \tilde\chi\(\o{t},t'\)\]_{\substack{ ji\\ lk}}.
\label{eq:abs_6}
\end{align}
Thus, the final equation for the longitudinal two-times linear response function becomes 
\begin{align}
\frac{d}{dt}\tilde\chi\(t,t'\)+\mathrm i\left[h_\mathrm{ext}\(t\),\tilde\chi\(t,t'\)\right]+\mathrm i\left[K^s\circ\chi(t,t') + \mathds{1} \gd\(t-t'\),\gr\(t\)\right]=-\int\,d\o{t}\,\[K ^\mathrm{dyn}\(t,\o{t}\)\circ \tilde\chi\(\o{t},t'\)\].
\label{eq:abs_8}
\end{align}
\end{widetext}
As explained in Sec.\ref{sec7} the non--local time dependence of the l.h.s. of Eq.\e{eq:abs_8} can be simplified by using the adiabatic ansatz. The most
complicate term remains the dynamical kernel, $K ^\mathrm{dyn}$, that we will extensively discuss in Sec.\ref{sec6E}. Indeed
$K ^\mathrm{dyn}$ is a common ingredient of the equations describing both the light absorption and emission.

\subsection{Light emission}
\label{sec6D}
Thanks to the quantization of the electromagnetic field we can now derive a closed expression for the light emission spectrum. As it will be clear shortly
the present formulation allows to introduce coherently the combined effects induced by the e--e and e--p coupling. This represents an important step forward
compared to the state--of--the--art formulation and, more importantly, it allows an efficient merging with DFT.

In order to derive the expression for the light emission spectrum in terms of the density matrix and use Eq.\e{eq:exp_2} to create a link with the
density--matrix equation of motion we use here the Poynting vector. In its hermitian form this vector is written as
\begin{multline}
\hat{\mathbf S}(1) = \frac{c}{8\pi}\[\hat{\mathbf E}^\dagger(1)\times\hat{\mathbf B}(1) - \hat{\mathbf B}^\dagger(1)\times\hat{\mathbf E}(1)\] \\
=\left.\frac{c}{8\pi}\[\hat{\mathbf E}^\dagger(1)\times\hat{\mathbf B}(2) - \hat{\mathbf B}^\dagger(2)\times\hat{\mathbf E}(1)\]\right|_{2=1}.
\label{eq:pl_1}
\end{multline}
The photo--emitted spectrum $I\(\go\)$ along the direction $\eta$ is obtained from the Fourier transformation of the macroscopic spatial average of the
projection of the Poynting vector along $\eta$,
$I\(\go\) \propto \la \hat{\mathbf S}\(\rr,\go\)\cdot \eta\ra$. 

We now take the relations between the fields and the vector potential 
\begin{gather}
\hat{\mathbf E}(1) = -\frac{1}{c}\frac{\partial \hat{\mathbf A}(1)}{\partial t_1},
\label{eq:pl_2}\\
\hat{\mathbf B}(1) = \nabla \times\hat{\mathbf A}(1).
\tag{\ref{eq:pl_2}$'$}
\end{gather}
By using Eq.\e{eq:pl_2} and Eq.(\ref{eq:pl_2}$'$) we can write explicitly the expression for the lesser and greater transverse photon propagator as
\begin{gather}
\mathcal D^<_{\ga\gb}(1,2) = \frac{1}{4\pi\mathrm i}\[\braket{\hat A_\gb(2)\hat A_\ga(1)} - \braket{\hat A_\ga(1)}\braket{\hat A_\gb(2)}\],
\label{eq:pl_3}\\
\mathcal D^>_{\ga\gb}(1,2) = \frac{1}{4\pi\mathrm i}\[\braket{\hat A_\ga(1)\hat A_\gb(2)} - \braket{\hat A_\ga(1)}\braket{\hat A_\gb(2)}\].
\tag{\ref{eq:pl_3}$'$}
\end{gather}
It is now possible to split the expectation value of the Poynting vector into two parts: a classical part which contains the information of the macroscopic effects
\begin{multline}
\braket{\hat{S}_\ga(1)}_\mathrm{class} = \frac{1}{2}\sum\limits_{
\substack{\gb=1 \\
\gb\neq \ga
}
}^3\frac{\partial\braket{\hat A_\gb(1)}}{\partial t_1}\times \\
\[\frac{\partial\braket{\hat A_\ga(2)}}{\partial r_{2,\gb}} - \frac{\partial\braket{\hat A_\gb(2)}}{\partial r_{2,\ga}}\],
\label{eq:s_class}
\end{multline}
and a contribution due solely to correlation effects
\begin{multline}
\braket{\hat{S}_\ga(1)}_\mathrm{corr} = \\\frac{1}{2}\sum\limits_{
\substack{\gb=1 \\
\gb\neq \ga
}
}^3\frac{\partial}{\partial t_1}\left\{
\frac{\partial}{\partial r_{2,\gb}}\[\mathcal D^>_{\gb\ga}(1,2) + \mathcal D^<_{\gb\ga}(1,2) \] - \right.\\
\left.\left.
\frac{\partial}{\partial r_{2,\ga}}\[\mathcal D^>_{\gb\gb}(1,2) + \mathcal D^<_{\gb\gb}(1,2)\]
\right\}\right|_{2=1}.
\label{eq:s_corr}
\end{multline}
Eq.\e{eq:s_corr} demonstrates that the evaluation of the light emission spectrum is linked to the calculation of the
lesser and greater transverse photon propagators that can be rewritten in terms of the advanced/retarded counterparts
and the transverse response function~\cite{hj-book}
\begin{align}
\overleftrightarrow{\mathcal D}^\gtrless (1,2) = \overleftrightarrow{\mathcal D}^r (1,3)
\overleftrightarrow{P}^\gtrless(3,4)\overleftrightarrow{\mathcal D}^a(4,2).
\label{eq:d_opt}
\end{align}
The problem of calculating the advanced and retarded photon propagators is itself a complicate issue. Here we assume to be interested in systems where the
renormalization of the photons can be neglected. This is the case of simple solids and molecules where the electronic polarisation effects on the
electromagnetic field can be assumed to be negligible. We start, then, from an  independent particle approximation for $\mathcal D^{a/r}$
\begin{gather}
\mathcal D^r_{\ga\gb}(\rr,t) = -\frac{\mathrm ic^2}{2}\sum_I\tau_{\ga\gb}^I\left[\xi_{I}(\rr)e^{-\mathrm i\omega_It} - h.c.\right]\theta(t),
\label{eq:d_r}\\
\mathcal D^a_{\ga\gb}(\rr,t) = \frac{\mathrm ic^2}{2}\sum_I\tau_{\ga\gb}^I\left[\xi_{I}(\rr)e^{-\mathrm i\omega_It} - h.c.\right]\theta(-t).
\tag{\ref{eq:d_r}'}
\end{gather}
In Eq.\e{eq:d_r} and Eq.(\ref{eq:d_r}$'$) we have introduced the single free photon wave--function, 
$\xi_{I}(\rr)\equiv 2\pi/\(\Omega \go_{\qq+\GG}\)e^{i\rr\cdot\(\qq+\GG\)}$. We remind here the convention introduced in Sec.\ref{sec5} to label the photon state
as $I\equiv\(\qq,\GG\)$.

By using Eq.\e{eq:d_r} and Eq.(\ref{eq:d_r}$'$) we rewrite Eq.\e{eq:s_corr} only in terms of $P_{\ga\gb}^\gtrless$
\begin{widetext}
\begin{multline}
\mathcal D_{\ga\gb}^>(1,2)+\mathcal D_{\ga\gb}^<(1,2)=\frac{c^4}{4}\sum_{\substack{I,J\\\ga_1\ga_2}}\int \mathrm d3 \mathrm d4\, \left[\xi_I(\rr_1-\rr_3)e^{-\mathrm i\omega_I(t_1-t_3)} - h.c.\right]\tau_{\ga\ga_1}^I\left[P_{\ga_1\ga_2}^>(3,4) + P_{\ga_1\ga_2}^<(3,4)\right]\tau_{\ga_2\gb}^J\times \\
\left[\xi_J(\rr_4-\rr_2)e^{-\mathrm i\omega_I(t_4-t_2)} - h.c.\right]\theta(t_1-t_3)\theta(t_2-t_4).
\label{eq:d_sum}
\end{multline}
\end{widetext}
Eq.\e{eq:d_sum} reduces the calculation of the light emission spectrum, $I\(\go\)$, to the evaluation of the equation of motion for 
$P_{\ga\gb}^\gtrless(1,2)$. It is evident, then, that we can now follow a path quite similar to the longitudinal case by using a series of chain rules to close
the equation of motion in an algebraic form. Indeed we start by the $P_{\ga\gb}^\gtrless(1,2)$ definition:
\begin{align}
P_{\ga\gb}(\rr_1,t_1;\rr_2,t_2)=-\frac{4\pi}{c}\frac{\delta\braket{J^\mathrm{ind}_\ga(\rr_1,t_1)}}{\delta\braket{A_\gb(\rr_2,t_2)}},
\label{eq:pol_1}
\end{align}
and by rewriting the induced current as
\begin{multline}
J_\ga^\mathrm{ind}(\rr,t) = \left.\mathrm i\Pi_\ga(\rr,\rr')G(\rr,t;\rr,t')\right|_{\substack{\rr'=\rr \\ t'=t^+}}=\\
=-\left.\Pi_\ga(\rr,\rr')\rho(\rr,\rr';t)\right|_{\rr'=\rr}.
\label{eq:j_rho}
\end{multline}
Then, by using Eq.~\e{eq:2nd_quant2} and expanding the density--matrix in the single--particle basis we can rewrite the induced current in terms of $\rho(t)$
\begin{align}
J_\ga^\mathrm{ind}(\rr,t) =-\sum_{ij}\Pi_{ij,\ga}(\rr)\rho_{ji}(t),
\label{eq:j_rho_t}
\end{align}
where $\Pi_{ij,\ga}(\rr) = \left. \Pi_{\ga}(\rr,\rr')\phi_i^*(\rr)\phi_j(\rr')\right|_{\rr'=\rr}$. We also expand the  variations of $\braket{A_\ga(1)}$ in the
photon basis
\begin{align}
\delta\braket{A_\ga(\rr,t)} = \frac{1}{2}\sum_I\left[\xi_I(\rr)\delta A_{I,\ga}(t) + h.c.\right].
\label{eq:d_A}
\end{align}
By using Eq.\e{eq:d_A} and Eq.\e{eq:j_rho_t} we can rewrite Eq.\e{eq:pol_1} as 
\begin{multline}
-\frac{4\pi}{c}\sum_{ij}\Pi_{ij,\ga}(\rr)\delta\rho_{ji}(t)=\\
\frac{1}{2}\sum_I\int \mathrm d t'\mathrm d\rr'\,P_{\ga\gb}(\rr,t;\rr',t')\left[\xi_I(\rr)\delta A_{I,\gb}(t) + h.c.\right].
\label{eq:delta_rho}
\end{multline}
We can now recast the functional derivative with respect to $\delta A$ as
\begin{align}
-\frac{4\pi}{c}\sum_{ij}\Pi_{ij,\ga}(\rr)\frac{\delta\rho_{ji}(t)}{\delta A_{I,\gb(t')}}=\frac{1}{2} \int\mathrm d\rr'P_{\ga\gb}(\rr, t;\rr', t')\xi_I(\rr'),
\label{eq:p_ia}
\end{align}
and its adjoint
\begin{multline}
-\frac{4\pi}{c}\sum_{ij}\Pi_{ij,\ga}(\rr)\frac{\delta\rho_{ji}(t)}{\delta A^*_{I,\gb(t')}}\\=\frac{1}{2} \[ \int\mathrm d\rr'P_{\ga\gb}(\rr, t;\rr',t')\xi_I(\rr')\]^*.
\end{multline}
At this point we can define the $I$--th component of the transverse linear response function as 
\begin{align}
 \mathbf P_{ij}^{I}(t,t') = \frac{\delta \gr_{ij}(t)}{\delta \mathbf A_{I}(t')}.
\label{eq:vec_chi}
\end{align}
This means that we can use the same procedure as we did for
the electronic polarisation function $\chi$ (see, for example, Eq.\e{eq:abs_3}) and write
\begin{multline}
\frac{\delta}{\delta \mathbf A_{I}(t')}\left[h_\mathrm{ext}(t),\rho(t)\right]_{ji} = \delta(t-t')\left[\mathbf p^{I},\rho(t)\right]_{ji} + \\ 
\left[h_\mathrm{ext}(t),\frac{\delta\rho(t)}{\delta \mathbf A_{I}(t')}\right]_{ji},
\label{eq:dh_da}
\end{multline}
with
\begin{align}
\mathbf p^{I}_{ij} = -\frac{\mathrm i}{c}\int \mathrm d\,\rr \phi_j^*(\rr)\xi_I(\rr)\nabla\phi_j(\rr).
\end{align}
In Eq.\e{eq:dh_da}  we have used the compact notation defined by Eq.~\eqref{eq:appA_4} in order to simplify the notation. The terms involving the functional derivatives of the COHSEX
and dynamical kernels will follow the same procedure of the longitudinal case. In the case of the COHSEX kernel contribution, indeed, we have that
\begin{multline}
\frac{\delta S_{ji} ^\mathrm{coh}(t)}{\delta \mathbf A_{I}\(t'\)}= \int\,\mathrm d\o{t} \,\frac{\delta\gr_{mn}\(\o{t}\)}{\delta \mathbf A_{I}\(t'\)} \frac{\delta
S_{ji} ^\mathrm{coh}(t)}{\delta\gr_{mn}\(\o{t}\)}=\\ =\left[K_s\circ\frac{\delta\gr\(t\)}{\delta \mathbf A_{I}\(t'\)},\gr(r)\right]_{ji},
\label{eq:dsc_da}
\end{multline}
and for the dynamical kernel we will write
\begin{multline}
\frac{\delta S_{ji} ^\mathrm{dyn}(t)}{\delta \mathbf A_{I}\(t'\)}= \int\,\mathrm d\o{t} \(K ^\mathrm{dyn}(t,\o t)\circ \frac{\delta\gr\(\o t\)}{\delta \mathbf A_{I}\(t'\)} \)_{ji}.
\label{eq:dsd_da}
\end{multline}
We can now derive the final equation of motion for $\mathbf P_{ij}^{I}(t,t')$ in similarity with what we did for $\chi(t,t')$
\begin{multline}
\frac{d}{dt} \mathbf P^{I}\(t,t'\)+\mathrm i\left[h_\mathrm{ext}\(t\), \mathbf P^{I}\(t,t'\)\right]+\\
\mathrm i\left[K_s\circ \mathbf P^{I}(t,t')+\delta(t-t')\mathbf p^{I},\gr(t)\right]=\\
- K ^\mathrm{dyn}\(t\)\circ \mathbf P^{I}\(t,t'\).
\label{eq:eom_vec_chi}
\end{multline}
Eq.\eqref{eq:eom_vec_chi} represents another important result of the present work. It derives a closed equation for the transverse response function and, in turns,
it provides a sound scheme to calculate the photo--luminescence spectrum. As this equation is derived in a scheme where e--e and e--p are included this means
that, physically, we have all essential ingredients of the dynamics.
The carriers excited by the primary pump pulse will, then, relax, dissipate and participate in bound electron--hole pairs before recombining and emitting light.

A final methodolgical remark is due. Eq.\e{eq:eom_vec_chi} is for the response function defined in Eq.~\eqref{eq:p_ia}. An equivalent equation can be
derived for its complex conjugate, defined by Eq.~(\ref{eq:p_ia}$'$). The only difference with Eq.\e{eq:eom_vec_chi} 
would be that the matrix elements $\mathbf p_{ij}^{I}$  must be replaced with $\mathbf q^{I}_{ij} = -\frac{\mathrm i}{c}\int \mathrm d\rr\,
\phi_j^*(\rr)\xi^*_I(\rr)\nabla\phi_j(\rr)$.

\subsection{The dynamical two--particle kernel}
\label{sec6E}
A common ingredient of Eq.\e{eq:abs_8a} and Eq.\e{eq:eom_vec_chi} is the dynamical kernel $K$, defined in Eq.\e{eq:abs_6} as 
$\frac{\delta S ^\mathrm{dyn}_{ji}\[\gr\]\(t\) }{\delta\gr_{mn}\(\o{t}\)}$. 
We have now all ingredients to analyze the physical contents of $K ^\mathrm{dyn}$ and interpret its properties. 

The problem is now how to calculate this functional derivative and how to use the LIA.

If we look back into Eq.~\eqref{eq:cca_16} and Fig.~\ref{fig:6} we see that how to evaluate the functional derivative of Eq.~\eqref{eq:abs_6}. The
point is what happens if we apply the LIA. Indeed we have two possibilities: to apply the LIA before doing the functional derivative; or 
to apply it after the derivative.
\begin{figure}[H]
\begin{center}
\epsfig{figure=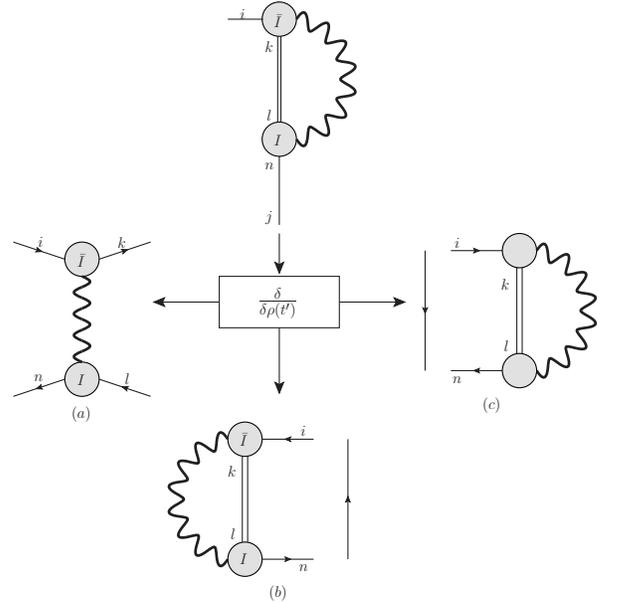,width=8cm}
\caption{\footnotesize{Diagrammatic contributions to the dynamical kernel $K ^\mathrm{dyn}$ resulting from the application of the functional derivative to $S ^\mathrm{dyn}$,
within the $GW$ approximation. }}
\label{fig:9}
\end{center}
\end{figure}

As we will show shortly this corresponds to neglect specific diagrams in the equation for the two--particle Green's function. In order to do so let us follow a
diagrammatic path. If we start from Fig.\ref{fig:6} we see that the derivative $\frac{\gd}{\delta\gr_{mn}\(\o{t}\)}$ can be applied both on $\gr$ 
and $\o{\gr}$. Graphically this corresponds to {\em open} the diagram as schematically shown in Fig.~\ref{fig:9}.

We clearly see that two kinds of interactions contribute to the dynamical kernel:
an electron--hole pair interaction (diagram (a) in Fig.~\ref{fig:9}), and a simple e--e (diagram (b)) or hole--hole (diagram (c)) interaction mediated by a
generic boson, shown in the bottom right of Fig.~\ref{fig:9}. The electron--hole interaction is, in the equilibrium language, the well--known dynamical part of
the screened electron--hole interaction. This has been studied in the framework of the BSE and showed to be connected and
compensated with the dynamical self--energy effects~\cite{PhysRevLett.91.176402,Bechstedt1997}.

Now, if we apply the LIA {\em before} doing the functional derivative we notice that the diagram (a) disappears as it comes from the internal density matrix
that, in Eq.\e{eq:exp_6} is approximated by its diagonal. Instead if we apply the LIA {\em after} the functional derivative the two internal
density matrices in diagrams\,(b) and (c) are approximated by occupation factors.

\begin{widetext}
If we take the path of applying the LIA {\em before} doing the functional derivative we get a closed expression for $K ^\mathrm{dyn}$:
\begin{align}
K_{\substack{ ji\\ lk}} ^\mathrm{dyn}\(t,\o{t}\)\approx K_{\substack{ ji\\ lk}} ^\mathrm{dyn}\(t\) \gd\(t-\o{t}\) =
\gd\(t-\o{t}\) \[\gd_{il}\(\gc^{\(-\)}_{ilk}\(t\)+\tilde{\gc}^{\(-\)}_{jlk}\(t\)\)+\gd_{jk}\((\gc^{\(+\)}_{lki}\(t\)+\tilde{\gc}^{\(+\)}_{lki}\(t\)\)\],
\label{eq:abs_7}
\end{align}
which allows us to simplify the equation of motion for the longitudinal two-times response function, turning Eq.~\e{eq:abs_8} into
\begin{align}
\frac{d}{dt}\tilde\chi\(t,t'\)+\mathrm i\left[h_\mathrm{ext}\(t\),\tilde\chi\(t,t'\)\right]+\mathrm i\left[K^s\circ\tilde\chi(t,t') + \mathds{1}
\gd\(t-t'\),\gr\(t\)\right]=-K ^\mathrm{dyn}\(t\)\circ \tilde\chi\(t,t'\).
\label{eq:abs_8a}
\end{align}
\end{widetext}
Eq.\e{eq:abs_8a} allows a simple and immediate physical interpretation as the $\gc$ function simply representing a time--dependent relaxation of the polarisation
that appears as a time--dependent broadening of the corresponding absorption peaks.

\section{The frequency representation of the transient absorption and luminescence spectra}
\label{sec7}
In the equilibrium limit it is well known that the response function depends on the time difference and therefore it can be easily transformed in frequency space
by applying a Fourier transform. This is a natural consequence of the time--translational invariance of the theory and reflects the fact the the energy of
the system is conserved. 

Out--of--equilibrium the energy of the electronic and nuclear sub--system is not conserved anymore as it flows back and forth to the electromagnetic
field. As a consequence both $\tilde\chi$ and $P$ are complex two--times functions.
The next step is to use the adiabatic ansatz (see~\ref{sec6A}) to change Eqs.~\eqref{eq:abs_8} and~\eqref{eq:eom_vec_chi} in order to have algebraic equations for $\tilde\chi$ and $P^{I,\ga}$, as is in the 
state--of--the--art equilibrium case. 

In the adiabatic ansatz, $\gr(t)$ and $A_{I,\ga}(t)$ will change slowly in time. For
times $(t,t')\approx \tau$ (with $\tau$ the pump--probe time delay) the response functions can be taken as a function of the relative time coordinate
\begin{gather}
\tilde\chi(t,t') \approx \tilde\chi^\tau(t-t'),\\
\mathbf P^I(t,t') \approx \mathbf P^{I,\tau}(t-t').
\end{gather}
Since in the adiabatic regime the density changes slowly in time, as long as the conditions described in section~\ref{sec6A} (a more detailed
description can be found in Ref.~\onlinecite{Perfetto2015}), the dynamic kernel of Eq.~\eqref{eq:abs_8} can also be taken as a function of the relative time coordinate. Thus,
we can finally rewrite Eq.~\eqref{eq:abs_8} into an algebraic form for the frequency dependent response function
\begin{multline}
-\mathrm i\go\tilde\chi^\tau(\go) + \mathrm i\[h_\mathrm{ext}(\tau),\tilde\chi^\tau(\go)\] + \\
\mathrm i\[K^{\tau}_s\circ\tilde\chi^\tau(\go)+1,\gr(\tau)\] = - K^\mathrm{dyn,\tau}(\go)\circ\tilde\chi^\tau(\go).
\label{eq:chi_go}
\end{multline}
The question now is how to bypass the calculation of the one particle density matrix $\gr(\tau)$ so that we can have an equation in which the only unknown
quantity is
the response function. In the equilibrium limit we can always rotate the hamiltonian $h_\mathrm{ext}$ and the density into a basis where both would be
diagonal, but this is not possible in non--equilibrium processes. Therefore, by following Ref.~\onlinecite{Perfetto2015}, we consider an orthogonal matrix $O(\tau)$ which rotates the hamiltonian to the
equilibrium basis and brings $h_\mathrm{ext}(\tau)$ to its diagonal form
\begin{align}
\[O^\dagger(\tau)h_\mathrm{ext}(\tau)O(\tau)\]_{ij} = \delta_{ij}\epsilon_i(\tau).
\end{align}
In this new basis the Eq.\e{eq:chi_go} reads
\begin{multline}
\[\go - \Delta \epsilon(\tau) +\mathrm iK^\mathrm{dyn,\tau}(\go)\]\circ\tilde\chi^\tau(\go) = \\
\[K^{\tau}_s\circ\tilde\chi(\omega) + 1,\gr(\tau)\].
\end{multline}
Here we have defined the energy tensor $\Delta \epsilon(\tau)_{\substack{ij\\pq}} = \[\epsilon_i(\tau)-\epsilon_j(\tau)\]1_{\substack{ij\\pq}}$. Following this
definition we can write the NEQ response function $\tilde\chi_0^\tau$  as
\begin{align}
\tilde\chi_0^\tau(\go) = -\[\go - \Delta\epsilon(\tau) + \mathrm iK^\mathrm{dyn,\tau}(\go)\]^{-1}\circ[\gr(\tau),1].
\label{eq:neq_x0}
\end{align}
If we now use the property 
\begin{align}
[\gr(\tau),K^\tau_s\circ\tilde\chi^\tau(\go)] = [\gr(\tau),1]\circ K^\tau_s\circ\tilde\chi^\tau(\go),
\label{eq:neq_x0_a}
\end{align}
we can finally obtain a Dyson-like equation for $\tilde\chi^\tau(\go)$
\begin{align}
\tilde\chi^\tau(\go) = \tilde\chi^\tau_0(\go) + \tilde\chi^\tau_0(\go)\circ K^\tau_s\circ \tilde\chi^\tau(\go).
\label{eq:dyson_chi}.
\end{align}
If we apply the same process to the transverse response function we see that its differential equation, Eq.~\eqref{eq:eom_vec_chi}, becomes
\begin{multline}
-\mathrm i\go \mathbf{P}^{I,\tau}(\go) + \mathrm i\[h_\mathrm{ext}(\tau),\mathbf{P}^{I,\tau}(\go)\] + \\
\mathrm i\[K^\tau_s\circ \mathbf{P}^{I,\tau}(\go) + \mathbf{p}^{I,\tau},\gr(\tau)\] = \\
- K^\mathrm{dyn,\tau}(\go)\circ \mathbf{P}^{I,\tau}(\go)
\end{multline}
and after reverting to the basis where $h_\mathrm{ext}(\tau)$ is diagonal and following the same procedure we used to arrive at Eq.~\eqref{eq:dyson_chi} we obtain
the desidered Dyson--like equation for ${P}^{I,\tau}(\go)$
\begin{multline}
\mathbf{P}^{I,\tau}(\go) = \mathbf{P}^{I,\tau}_0(\go)+ \mathbf{P}^{I,\tau}_0(\go)\circ K^\tau_s\circ \mathbf{P}^{I,\tau}(\go)
\label{eq:dyson_p}
\end{multline}
with the NEQ $\mathbf{P}^{I,\tau}_0(\go)$ being
\begin{multline}
\mathbf{P}^{I,\tau}_0(\go) = \\ -\[\go - \Delta\epsilon(\tau) + \mathrm iK^\mathrm{dyn,\tau}(\go)\]^{-1}\circ[\gr(\tau),\mathbf{p}^{I,\tau}].
\label{eq:neq_p0}
\end{multline}
Eq.\e{eq:neq_p0} represents another crucial result of the present work. It provides, indeed, the basis for a fully {\em ab--initio}
implementation of the transient photo--luminescence spectrum. This will allow to extend what has been already done in previously~\cite{MoS2_pogna} in the transient absorption case.

\section{The merging with Density--Functional Theory}
\label{sec8}
In order to merge the BKE in the general DFT scheme we follow the same strategy used in the standard MBPT approach~\cite{Onida2002}. This is based on the
use of the Kohn--Sham\,(KS) Hamiltonian, $h_{KS}$ as the reference single particle Hamiltonian. This means, in practice, that in Eq.\e{sec2:eq:full_h} the $h$
operator is replaced by
\begin{multline}
h_{KS}\phi_i(\rr)=\\
=\[-\frac{\nabla^2}{2} + v_\mathrm{ext}\(\rr\) + v_H\[\gr\]\(\rr\) + v_{xc}\[\gr\]\(\rr\)\]\phi_i(\rr)= \\
= \epsilon_i^{KS}\phi_i(\rr),
\label{eq:ks_eq}
\end{multline}
with $v_H$ and $v_{xc}$ the Hartree and the exchange--correlation potentials which, in DFT, are functional of the electronic density $\gr\(\rr\)$. 
Then, $\phi_i$ represents the basis of wave functions used in Eqs.~\eqref{eq:cca_1} and~\eqref{eq:cca_2}.

DFT is a mean-field theory, where the electronic system is described by a group of pseudo non--interacting particles which move under the influence of the
$v_{xc}$ potential that already contains some of the e--e correlation effects. This initial correlation already present in the KS Hamiltonian has
several important consequences in the electronic~\cite{GW_review} and also in the phononic dynamics\cite{PhysRevB.91.224310}. 

As far as the electronic dynamics is concerned $v_{xc}$ can lead to subtle double--counting problems that are safely removed by defining, in
Eq.\e{sec3:eq:eom_ho},
\begin{equation}
h_\mathrm{ext} = h_{KS} - v_{xc}.
\label{eq:5.2}
\end{equation}
Another important ingredient of the present approach are the phonon modes. These, whithin the DFT scheme, are obtained by considering the total Hamiltonian $H$
as a functional of the atomic positions $\{\RR\}$. The problem of finding the phonon modes reduces to the self--consistent calculation of derivatives of $H$.
Indeed, if DFT is a self--consistent theory,
DFPT~\cite{baroni2001,Gonze1995} is its extension to take into account, self--consistently, the effect of static perturbations (like nuclear displacements). In this
case, DPFT provides an exact description of phonons within the limits of a static and adiabatic approach. 

Thus DFT and DFPT provides all ingredients of the present theory and allow a full {\em ab--initio} implementation. Indeed, as discussed in Sec.\ref{sec5}, we
have that for each element of the theory we can define a DFT/DFPT counterpart.

(a) As far as the single--particle electronic basis is concerned the  KS wave--functions $\phi_i$ represent a natural definition. In this basis all standard
MBPT machinery can be used to calculate the ingredients of standard Hedin's equations~\cite{Onida2002}.

(b) In the e--p case phonons and e--p interaction matrix elements can be easily calculated within DFTP. Indeed, in this case, the
ionic potential $V_n$ appearing in Eq.\e{eq:cca_7} is
\begin{equation}
V_n\(\rr\) \Rightarrow V_{scf}\(\RR,\rr\)= v_{H}\(\rr\)+v_{xc}\(\rr\)-\sum_{\RR}\frac{Z_{\RR}}{|\rr-\RR|},
\label{eq:5.3}
\end{equation}
and the phonon frequencies can be safely calculated within DFPT.

These simple connections demostrates that, by simply implementing the NEGF framework in a KS basis, we can extend the predictive power of the many--body
technique beyond the standard and well--known equilibrium limit. 

\section{Conclusions}
\label{conclusions}
In this work we have carefully introduced, step by step, all mathematical and procedural steps needed to derive a simplified but 
accurate approach to the  theory of out--of--equilibrium systems. We have started by the conclusions of previous 
works to integrate the electron--electron,
electron--phonon and electron--photon interactions in a consistent scheme based on the non--equilibrium Green's function theory.

We first derive a very general set of Hedin's equations where phonons and photons appear as the longitudinal and transverse components of the theory. We
introduce, then, the correspondinf longitudinal and transverse response and vertex functions showing how they are deeply connected. 

However the complete theory appears clearly to be, by far, too complex to be implemented in realistic materials like nano--structures, surfaces and all kind
of systems currently within the range of applicability of DFT. As a consequence, in order to provide a workable scheme that can be potentially implemented in a
DFT basis, we adopt the well--known $GW$ approximation that is extended to the out--of--equilibrium regime and to the transverse component of the
self--energy. 

This simplified
form allows us to introduce a series of approximations: the Generalized Baym--Kadanoff ansatz, the Completed Collision approximation, the Low--Intensity
approximation and the adiabatic ansatz. We discuss the physical motivations beyond these approximations and their range of applicability. By concentrating on
the slow and weakly perturbed components of the dynamics we derive a set of closed equations for the carrier dynamics, transient absorption and
transient light--emission written solely in terms of the time--dependent density--matrix.

This represents a drammatic simplification of the scheme that can, now, be applied in a DFT framework to realistic materials.
This is, indeed, the main conclusion of this work. The advanced and involved NEGF theory can be made simple enough to be applied in an \ai\,framework and
provide the basis for a new set of tools available to the scientific community that we can group within the scopes a newly defined {\em ai}--NEGF approach.

The {\em ai}--NEGF extends the predictive power of the {\em ai}--MBPT approach to describe {\em realistic systems} out--of--equilibrium thus
opening the path to new applications, basic research and numerical and methodological developments. This work represents, indeed, the formal proposal for such a
working approach opening the path to many future developments and applications.

\section*{Acknowledgements}
We acknowledge financial support by the {\em Futuro in Ricerca} grant No. RBFR12SW0J of the Italian Ministry of Education, University and Research MIUR. AM also
acknowledges the funding received from the European Union project MaX {\em Materials design at the eXascale} H2020-EINFRA-2015-1, Grant agreement n. 676598 and
{\em Nanoscience Foundries and Fine Analysis - Europe} H2020-INFRAIA-2014-2015, Grant agreement n. 654360. PM thanks the Portuguese Foundation for Science and
Technology (FCT) and the Portuguese Ministry for Education and Science, for the funding received through the scholarship SFRH/BD/84032/2012 under the European Social Fund and the Programa Operacional Potencial Humano (POPH).

\appendix
%
%
%
%
%

\section{Conventions for linear algebra operations}
\label{appA}
In the derivation of the equations of the section~\ref{sec7} we introduced the following {\em ad--hoc} notation for the generic products of matrices
\begin{gather}
(M\circ V)_{pq}\equiv M_{\substack{ pq\\mn}}V_{nm},
\label{eq:appA_1}\\
(T\circ M \circ V) \equiv T_{pq} M_{\substack{ qp\\ mn}}V_{nm}, 
\label{eq:appA_2}\\
(M \circ N)_{\substack{ mn\\ rs}}\equiv M_{\substack{ mn\\pq}} 
N_{\substack{ qp\\rs}},
\label{eq:appA_3}\\
[N , V]_{\substack{ mn\\ pq}}=-[V , N]_{\substack{ mn\\ pq}}\equiv N_{\substack{ mi\\pq}} V_{in} - V_{mi} 
N_{\substack{ in\\pq}}.
\label{eq:appA_4}\\
[N , V]_{mn}=-[V , N]_{mn}\equiv N_{mi} V_{in} - V_{mi}N_{in}.
\label{eq:appA_5}
\end{gather}

\bibliographystyle{apsrev4-1}
\bibliography{refs}	

\end{document}

%% file: alias.tex
\newcommand{\m}{\mathrm}

\newcommand{\be}{\begin{equation}}
\newcommand{\ee}{\end{equation}}
\newcommand{\bea}{\begin{eqnarray}}
\newcommand{\eea}{\end{eqnarray}}

\newcommand{\e}[1]{{\eqref{#1}}}
\renewcommand{\o}[1]{{\overline{#1}}}

\def\T		{\left(T\right)}

\def\rr		{{\bf r}}

\def\AA		{{\bf A}}
\def\dd		{{\bf d}}

\def\GG		{{\bf G}}

\def\RR		{{\bf R}}

\def\EE		{{\bf E}}
\def\qq		{{\bf q}}

\def\JJ		{{\bf J}}
\def\jj		{{\bf j}}

\def\ga         {\alpha}
\def\gb         {\beta}
\def\gc         {\gamma}
\def\gs         {\sigma}

\def\gd         {\delta}

\def\gee        {\epsilon}
\def\gl         {\lambda}

\def\go         {\omega}

\def\gr         {\rho}
\def\gs         {\sigma}

\def\gt         {\theta}

\def\capo       {\right.\\ \left.}

\def\la         {\langle}
\def\ra         {\rangle}

\renewcommand{\[}{\left[}
\renewcommand{\]}{\right]}
\renewcommand{\(}{\left(}
\renewcommand{\)}{\right)}

\def\bverb      {\renewcommand{\baselinestretch}{1}\begin{verbatim}}
\def\everb  	{\end{verbatim}\renewcommand{\baselinestretch}{1.3}}

\def\ai         {{\it ab-initio}}
\def\ga         {\alpha}
\def\gb         {\beta}

\def\gee        {\varepsilon}
\def\gl         {\lambda}

\def\go         {\omega}

\def\la         {\langle}
\def\ra         {\rangle}

\def\qq         {{\mathbf q}}
\def\rr         {{\mathbf r}}

\renewcommand{\[}{\left[}
\renewcommand{\]}{\right]}
\renewcommand{\(}{\left(}
\renewcommand{\)}{\right)}

\newcommand{\cnr} {Istituto di Struttura della Materia of the National Research Council, Via Salaria Km 29.3,I-00016 Monterotondo Stazione, Italy}
\newcommand{\etsf} {European Theoretical Spectroscopy Facilities (ETSF)}
\newcommand{\cfc} {CFisUC, Department of Physics, University of Coimbra, Rua Larga, 3004-516 Coimbra, Portugal}